\documentclass[superscriptaddress,twocolumn,floatf  ix,nofootinbib,a4paper,aps,reprint]{revtex4-1}

\usepackage{amsmath}
\usepackage{microtype}
\usepackage{graphicx}
    \graphicspath{{figs/}}
\usepackage[caption=false]{subfig}
\usepackage{overpic}
\usepackage{booktabs} 

\usepackage{bm}
\usepackage[hidelinks]{hyperref}

\usepackage{algorithm}
\usepackage{algorithmic}

\usepackage{tikz}
\usetikzlibrary{calc}
\newcommand{\tikzmark}[1]{\tikz[overlay,remember picture,baseline] \node [anchor=base] (#1) {};}%

\def\drawbox{%
\begin{tikzpicture}[remember picture, overlay]
\draw[black] ($(left|-begin)+(0.9cm,-2pt)$) rectangle ($(left|-end)+(5.3cm,-2pt)$);
\end{tikzpicture}%
}

\usepackage{xcolor}
\definecolor{purple}{RGB}{128,0,128}

\usepackage[normalem]{ulem}

\begin{document}

\title{Efficient training of energy-based models via spin-glass control}

\author{Alejandro Pozas-Kerstjens}
\thanks{These authors contributed equally to the work}
\affiliation{Departamento de An\'alisis Matem\'atico, Universidad Complutense de Madrid, 28040 Madrid, Spain}
\altaffiliation{Correspondence should be addressed to \href{mailto:physics@alexpozas.com}{physics@alexpozas.com}, \href{mailto:gorka.munoz@icfo.eu}{gorka.munoz@icfo.eu}, \href{mailto:grzyb@amu.edu.pl}{grzyb@amu.edu.pl}}

\author{Gorka Mu\~noz-Gil}
\thanks{These authors contributed equally to the work}
\affiliation{ICFO-Institut de Ci\`encies Fot\`oniques, The Barcelona Institute of Science and Technology, 08860 Castelldefels (Barcelona), Spain}
\altaffiliation{Correspondence should be addressed to \href{mailto:physics@alexpozas.com}{physics@alexpozas.com}, \href{mailto:gorka.munoz@icfo.eu}{gorka.munoz@icfo.eu}, \href{mailto:grzyb@amu.edu.pl}{grzyb@amu.edu.pl}}

\author{Eloy Pi\~nol}
\affiliation{ICFO-Institut de Ci\`encies Fot\`oniques, The Barcelona Institute of Science and Technology, 08860 Castelldefels (Barcelona), Spain}
\affiliation{Instituto Universitario de Matem\'atica Pura y Aplicada, Universitat Polit\`ecnica de Val\`encia, 46022 Valencia, Spain}

\author{Miguel \'Angel Garc\'ia-March}
\affiliation{Instituto Universitario de Matem\'atica Pura y Aplicada, Universitat Polit\`ecnica de Val\`encia, 46022 Valencia, Spain}

\author{Antonio Ac\'in}
\affiliation{ICFO-Institut de Ci\`encies Fot\`oniques, The Barcelona Institute of Science and Technology, 08860 Castelldefels (Barcelona), Spain}
\affiliation{ICREA, Passeig Lluis Companys 23, 08010 Barcelona, Spain}

\author{Maciej Lewenstein}
\affiliation{ICFO-Institut de Ci\`encies Fot\`oniques, The Barcelona Institute of Science and Technology, 08860 Castelldefels (Barcelona), Spain}
\affiliation{ICREA, Passeig Lluis Companys 23, 08010 Barcelona, Spain}

\author{Przemys{\l}aw R. Grzybowski}
\affiliation{Faculty of Physics, Adam Mickiewicz University, Umultowska 85, 61-614 Pozna{\'n}, Poland}
\altaffiliation{Correspondence should be addressed to \href{mailto:physics@alexpozas.com}{physics@alexpozas.com}, \href{mailto:gorka.munoz@icfo.eu}{gorka.munoz@icfo.eu}, \href{mailto:grzyb@amu.edu.pl}{grzyb@amu.edu.pl}}

\begin{abstract}
We introduce a new family of energy-based probabilistic graphical models for efficient unsupervised learning.
Its definition is motivated by the control of the spin-glass properties of the Ising model described by the weights of Boltzmann machines.
We use it to learn the Bars and Stripes dataset of various sizes and the MNIST dataset, and show how they quickly achieve the performance offered by standard methods for unsupervised learning.
Our results indicate that the standard initialization of Boltzmann machines with random weights equivalent to spin-glass models is an unnecessary bottleneck in the process of training.
Furthermore, this new family allows for very easy access to low-energy configurations, which points to new, efficient training algorithms.
The simplest variant of such algorithms approximates the negative phase of the log-likelihood gradient with no Markov chain Monte Carlo sampling costs at all, and with an accuracy sufficient to achieve good learning and generalization.

\end{abstract}

\maketitle

\section{Introduction}
Machine learning has emerged as a disruptive technology transforming industries, society and science.
Its perhaps most remarkable recent developments are based on supervised and reinforcement learning in deep neural networks.
Yet unsupervised learning is expected to be much more important in the long term~\cite{2015LecunNature,Du2019Review}.
Energy-based models, with their ability of unsupervised learning of probability distributions for generative purposes, are promising building blocks of future machine learning systems.
Among them, Boltzmann machines (BMs) have especially prospective properties: their latent variables allow for deep neural network architectures while the learning algorithm is remarkably simple~\cite{1983HintonProceedings,2002Hinton,2005Mnih}.

Training BMs is nevertheless hard due to the need of obtaining samples from the models built.
Specifically, a set of averages with respect to training data and the defined model needs to be determined at every learning step.
In general, such averages cannot be computed exactly for large networks because of the large dimension of the vector spaces involved.
Instead, they are estimated, for instance, by sampling through Markov chain Monte Carlo (MCMC) methods.
Initial sampling heuristics relied on short-step Gibbs or Metropolis-Hastings methods, which were soon complemented with features such as persistent chains~\cite{Tieleman2008pcd} or with replicas of the original chains~\cite{Hukushima2003Population}.
These improvements come, however, associated with increased memory and computational costs.
Given that energy-based models are closely related to problems of statistical physics, the powerful methods developed for statistical physics are among the most promising for dealing with the problem of training BMs.
These include modern MCMC algorithms for physical systems like Parallel Tempering~\cite{Desjardins2010pt} or Simulated Annealing~\cite{1992MarinariEPL}.
The problem of training BMs is so relevant and challenging that special hardware systems exploiting specific physical processes have been developed to deal with the task of sampling.
These include systems operating in the regime of classical physics~\cite{Inagaki2016NTT,McMahon2016NTT}, as well as based on purely quantum or hybrid classical-quantum machines~\cite{2018AminPRX,2018Perdomo_OrtizQST}.
While these routes are promising, they have important drawbacks when faced with practical applications, mostly due to the immature state of these novel computing platforms.

The problem of training BMs can be framed in the context of statistical physics and benefit from its associated theoretical body.
Indeed, the connection between BMs and statistical mechanics is known since the initial developments in the field~\cite{1983HintonProceedings}.
From this point of view, neurons in BMs play the role of physical spins of an Ising model, the weights represent the coupling strengths between spins, and the biases of the neurons are local fields affecting each individual spin.
Once set this analogy, it is natural to identify the BM initialized with independently drawn random weights with the Sherrington-Kirkpatrick spin-glass (SKSG) model~\cite{SpinGlassRev}.
Thus, the difficulty of training BMs through sampling is connected to the difficulty of determining the ground state energy of the SKSG model on non-planar graphs, which is an NP-complete problem.

In this work we find that the typical initialization of BMs with random weights equivalent to the SKSG model is an unnecessary bottleneck in the process of training.
We consequently propose a radically different approach: we regularize the couplings in the Boltzmann machine in order to avoid a spin-glass behavior at any point of training.
Thus, this indicates an alternative to pursuing the paramount problem of efficient sampling in the SKSG model.
We call this method Regularized Axons (RA), and the family of models that it gives rise to, RA-BMs.
Moreover, RA provides proxies of low-energy configurations, which suggest new methods for estimating the gradient of the log-likelihood function that is optimized during training.
In particular, we show a simple case where MCMC sampling is not necessary for successfully learning a dataset.
This method, which we term training via Pattern-InDuced correlations (PID), thus reduces the numerical effort of training to a minimum.
Although in this work the numerical examples focus on restricted BMs (RBMs), the main ideas remain applicable to any energy-based model with an energy function similar to an Ising model with random weights and, in particular, deep BMs.

We first show in a conventional academic example that during training of standard RBMs two main phenomena occur: on one hand, the ability to access low-energy states rises dramatically, and on the other, the models' weights evolve in such a way that standard RBM models resemble RA-RBMs after training.
These phenomena signal essential differences between a well-trained model and the SKSG model.
Then, we show that avoiding the spin-glass regime during training via RA allows to obtain well-trained models.
We do this by demonstrating on several examples of increasing complexity that models with RA are capable of fast and successful learning and generalization, where in some instances PID contributes by reducing further the numerical effort.
With this, we conclude that the regularization we impose is not restrictive when it comes to the expressive power of the model.

This manuscript is organized as follows: after a short introduction to the formalism of Boltzmann machines in Section~\ref{sec:intro}, in Section~\ref{sec:method} we describe the technical results of our work: RA for regularizing BM models, and PID for training them.
Section~\ref{sec:rationale} is devoted to their justification, based on arguments coming from the theory of statistical physics.
In Section~\ref{sec:results} we empirically test the performance of RAPID in various datasets, showing its efficient learning and its generalization ability.
We conclude with a discussion and point out relevant remarks in Section~\ref{sec:conclusions}.

\section{Preliminaries: Boltzmann machines}\label{sec:intro}
We begin by recalling the standard BM, which consists of $N$ binary neurons $\bm\sigma$ (here we use values $\sigma_j=\pm 1$, which are standard in the physics of spin systems), separated into two disjoint sets of $V$ visible and $H$ hidden neurons, which will be referred to respectively as $\bm v$ and $\bm h$, so that $\bm\sigma=(\bm v,\bm h)$. The energy of a given configuration of neurons is defined as:
\begin{equation}
\label{eq:energy}
E(\bm{\sigma})=-\sum_{ij}^NW_{ij}\sigma_i\sigma_j-\sum_{i}^Nb_{i}\sigma_i,
\end{equation}
where the weights $W_{ij}$ describe connections (axons) between neurons, while $b_i$ are local biases. Alternatively, such BM setup describes spin systems where the weights describe interactions between pairs of spins and the biases are local magnetic fields.
Different architectures of connections (i.e. different graphs whose vertices are neurons and edges denote non-zero weights) can be considered.
For example, in RBMs, there are only connections between visible and hidden neurons, and all visible-visible and hidden-hidden connections are set to zero.
However, in the most general case the neural network is fully connected.
In the following, and throughout the whole manuscript, we will neglect biases, as the main issues we discuss are related to the distribution of weights.

The probability of a model having a visible configuration $ \bm{v}$, $P_{\rm{model}}( \bm{v})$ , is given by a Boltzmann distribution
\begin{equation}
\label{eq:boltz_prob}
P_{\rm{model}}(\bm{v})=\frac{\sum_{\bm h} e^{-E(\bm{v},\bm{h})}}{\sum_{\bm \sigma}e^{-E(\bm{\sigma})}}.
\end{equation}
The goal of the training is to determine the parameters $W_{ij}$ of the energy function~\eqref{eq:energy} such that $P_{\rm{model}}(\bm{v})$ represents as close as possible the distribution $P_\mathrm{data}$ underlying some training dataset $\mathcal{T}$. This is usually done by minimizing the negative log-likelihood (NLL),
\begin{equation}
\label{eq:nll}
\mathcal{L}=-\sum_{\bm v\in\mathcal{T}} P_\mathrm{data}(\bm{v})\log P_\mathrm{model}(\bm{v}),
\end{equation}
with respect to the parameters of the energy function.
Let us collectively denote these parameters by $\theta$.
As $P_\mathrm{data}$ is independent on these parameters, the minimization is only performed to $\log P_\mathrm{model}$.
The derivative of this term takes the form of
\begin{equation}
\label{eq:deriv_BM}
    \partial_\theta (-\log P_\mathrm{model})=\left< \partial_\theta E \right>_\mathrm{data}-\left< \partial_\theta E \right>_\mathrm{model},
\end{equation}
where the bracket $\left< \cdot \right>$ denotes the expectation value with respect to the probability distributions $P_\mathrm{data}$ or $P_\mathrm{model}$ for the data and model averages, respectively.
Sampling from such distributions is the main challenge of BMs, as discussed previously.
In fact, RBMs were introduced in order to facilitate the computation of $\left< \cdot \right>_\mathrm{data}$~\cite{smolensky}.
However, even for RBMs, the computation of $\left< \cdot \right>_\mathrm{model}$ is still very difficult if the weights are random.

\section{RAPID---Regularized Axons and Pattern-InDuced correlations}\label{sec:method}
This section contains our main technical contribution, the definition of a family of energy-based probabilistic graphical models that avoids the training difficulties that stem from spin-glass phenomenology.
This family, which we call Boltzmann machines with Regularized Axons, or RA-BMs, is introduced in Section~\ref{sec:method:ra}.
The procedure of regularizing the Ising model couplings (i.e., the BM weights) defines a simple form of the space of configurations with low energy, which can be used for approximating averages under the model distribution in a very resource-efficient manner.
We employ such property in Section~\ref{sec:method:pid} to define an algorithm for training via Pattern-InDuced correlations (PID).

\subsection{Regularized Axons}\label{sec:method:ra}
We employ a regularization of the weights of the BM by constructing them from a number $K$ of configurations called \textit{patterns}, each described by a set of variables $\{\bm\xi^{(k)}\}_{k=1}^K$ where $\xi_i^{(k)}\in \{-1,+1\} \ \forall \ k=1\dots K$, $i=1\dots N$:
\begin{equation}
\label{eq:hebbian_rule}
W_{ij}=\frac{1}{\sqrt{K}}\sum_{k=1}^K \xi_i^{(k)}\xi_j^{(k)}. \end{equation}
Note that, with this form, the weights are naturally constrained to lie in the interval $[-\sqrt{K},\sqrt{K}]$.
Such form of the weights is well known in machine learning from the Hopfield model of associative memory~\cite{1974little,1982Hopfield}, which implements the Hebbian rule so that ``neurons wire together if they fire together''~\cite{1949Hebb}.
Contrarily to the original Hopfield model, our patterns do \textit{not} represent memorized data. We discuss in detail the differences between the Hopfield model and our proposal in Section~\ref{sec:pattern:analysis}. Here, we just note that \textit{the patterns are the trainable parameters} of the model.
For BMs with restricted connectivity like RBMs or deep BMs, one should notice that some $W_{ij}$ will be set to $0$ and not calculated according to Eq.~\eqref{eq:hebbian_rule}.

Importantly, if one considers $K\,{\ll}\, N$, then the patterns are \textit{explicit low-energy configurations} of the Ising model associated to the neural network with weights given by Eq.~\eqref{eq:hebbian_rule}~\cite{PhysRevA.32.1007}.
Furthermore, the condition $K\ll N$ ensures that at low temperatures the model is not in the spin-glass phase~\cite{1985Amit,amit_1989}, which is the primary motivation for such regularization.
We refer the reader to Section~\ref{sec:rationale_reg} for more details on these statements.
Therefore, in a typical training instance of an RA-BM, one would proceed to first choose a number of patterns $K$ high enough to faithfully learn the data (this is, to ensure that the model has enough \textit{plasticity}), and only then choose the number of hidden neurons in such a way that $K\,{\ll}\, N$.

\subsection{Training via Pattern-InDuced correlations}
\label{sec:method:pid}
For weights regularized via Eq.~\eqref{eq:hebbian_rule}, the patterns $\{\bm\xi^{(k)}\}_k$ are themselves low-energy configurations of the spin model of Eq.~\eqref{eq:energy} when $K\,{\ll}\, N$.
Recalling that Boltzmann distributions of the form~\eqref{eq:boltz_prob} give exponentially larger weights to low-energy configurations, averages under the model distribution, and in particular the negative phase of Eq.~\eqref{eq:deriv_BM}, can be well approximated by the corresponding averages over the values of the spins in the patterns.
This is,
\begin{equation}
    \left\langle f(\bm\sigma)\right\rangle_\mathrm{model}\overset{\text{PID}}{\approx}\frac{1}{K}\sum_{k=1}^K f(\bm\xi^{(k)}),
    \label{eq:pid}
\end{equation}
where $f(\bm\sigma)$ is an arbitrary function of the neurons in the model.
We refer to this procedure as estimation through Pattern-InDuced correlations, or PID.

As training progresses, the patterns $\{\bm\xi^{(k)}\}_k$ can acquire non-trivial overlaps with each other, losing the guarantee that they represent an exhaustive set of low-energy configurations of the Ising model associated to the BM.
Importantly, due to their initial construction, such patterns still lie close to \textit{different} energy minima each.
This ensures a fair calculation of averages, and implies that the patterns serve as ideal seeds for iterations of Gibbs sampling.
In Section~\ref{sec:results} we show via examples how RA-RBMs trained with PID without Gibbs sampling are capable of learning simple datasets, while as few as a single Gibbs step is enough to learn complex ones.

The algorithmic form of RAPID, the training of an RA-BM via PID, is presented in Algorithm~\ref{alg:rapid} for the particular case of an RBM architecture.
The highlighted step is the calculation of the negative phase by means of PID, and the remaining is common to any RA-RBM.
The general-case algorithm for arbitrary, deep or fully connected BMs, can be straightforwardly obtained from Algorithm~\ref{alg:rapid}.
\begin{algorithm}[H]
  \caption{Learn dataset with an RA-RBM and PID}
  \label{alg:rapid}
  \tikzmark{left}
  \begin{algorithmic}
    \STATE {\bfseries Input:} dataset $\mathcal{X}=\{\bm v^{(i)}\}_i$,
    \STATE $\qquad\quad\,\,\,$number of patterns $K$,
    \STATE $\qquad\quad\,\,\,$hidden layer size $H$ s.t. $K\,{\ll}\, H+\mathrm{length}(\bm v^{(i)})$,
    \STATE $\qquad\quad\,\,\,$learning rate $\lambda$, number of epochs $E$
    \STATE $V\leftarrow\mathrm{length}(\bm v^{(1)})$
    \FOR{$k=1$ {\bfseries to} $K$}
    \STATE Initialize $\bm\xi^{(k)}_v\in\{-1,+1\}^{V}$ randomly
    \STATE Initialize $\bm\xi^{(k)}_h\in\{-1,+1\}^{H}$ randomly
    \STATE $\bm\xi^{(k)}\leftarrow \mathrm{concatenate}(\bm\xi^{(k)}_v,\bm\xi^{(k)}_h)$
    \ENDFOR
    \STATE $W_{ij}\leftarrow\frac{1}{\sqrt{K}}\sum_{k=1}^K\xi_i^{(k)}\xi_{V+j}^{(k)}$
    \FOR{$e=1$ {\bfseries to} $E$}
    \FOR{$\bm v$ {\bfseries in} $\mathcal{X}$}
    \STATE $\bm h\leftarrow\mathrm{get}\_\mathrm{h}\_\mathrm{from}\_\mathrm{v}(\bm v, W)$
    \STATE $\bm p^{(k)}\leftarrow\mathrm{get}\_\mathrm{phase}(\bm v,\bm h,\bm\xi^{(k)})$
    \tikzmark{begin}\STATE $\bm n^{(k)}\leftarrow\mathrm{get}\_\mathrm{phase}(\bm\xi_v,\bm\xi_h,\bm\xi^{(k)})$\tikzmark{end}
    \STATE $\bm\xi^{(k)}\leftarrow\bm\xi^{(k)}+\lambda(\bm p^{(k)}-\bm n^{(k)})$
    \STATE $W_{ij}\leftarrow\frac{1}{\sqrt{K}}\sum_{k=1}^K\xi_i^{(k)}\xi_{V+j}^{(k)}$
    \ENDFOR
    \STATE $\bm\xi^{(k)}\leftarrow\mathrm{binarize}(\bm\xi^{(k)})$
    \STATE $W_{ij}\leftarrow\frac{1}{\sqrt{K}}\sum_{k=1}^K\xi_i^{(k)}\xi_{V+j}^{(k)}$
    \ENDFOR
  \end{algorithmic}
  \drawbox
\end{algorithm}
Note that the function \mbox{$\mathrm{get}\_\mathrm{phase}()$} is typically the average of the gradient of the free energy. In Appendix~\ref{app:updaterule} we give its explicit form for an RA-RBM.

An important aspect to notice is that, after an update, the parameters $\bm \xi^{(k)}$ depart from taking values from  $\{-1,+1\}^N$.
Thus, they do not represent exactly spin configurations, although they usually remain close to $\pm1$.
In order to solve this problem, we binarize the parameters back after each epoch of training (see the third-to-last line of Algorithm~\ref{alg:rapid}).
Different procedures, such as those we propose in Appendix~\ref{app:discrete} and use in the experimental analysis of Section~\ref{sec:results}, can be employed.
Also, it must be noted that RA and PID are independent results and, in particular, it is possible to replace PID with other techniques for approximating the negative phase of the parameter updates.

In summary, the novelty of the combination of Regularized Axons and training via Pattern-InDuced correlations, RAPID, comes from: (i) avoiding the SKSG phase at any moment of training by utilizing weights constructed via Eq.~\eqref{eq:hebbian_rule} while scaling $H$ to keep $K\ll N$; and (ii) exploiting the patterns introduced in Eq.~\eqref{eq:hebbian_rule} for approximating the low-energy space of the associated spin model in an efficient way and using them to approximate the negative phase.
As we show in Section~\ref{sec:results}, this recipe is sufficient for employing RBMs to learn relevant probability distributions.

\section{Physical explanation}\label{sec:rationale}
In this section we explain the theoretical justification for RA-BMs, which originates in the field of statistical physics.

\subsection{Hardness of sampling and spin glasses}
Perhaps the most profound result stemming from the perspective of statistical physics in BMs is the understanding of the origin of the hardness of sampling the models.
The Boltzmann probability distribution, Eq.~\eqref{eq:boltz_prob}, is dominated by contributions from low-energy configurations, and a good sampling technique must probe such configurations well.
However, determining the lowest-energy configuration---also known as \textit{ground state}---of any Ising model defined on a non-planar graph with independently drawn couplings is an NP-complete problem~\cite{Barahona1982complexity}.
An example of such models is the usual starting point of a BM.
At the beginning of training, when typically the couplings between neurons are drawn at random, a BM is equivalent to the Sherrington-Kirkpatrick spin-glass model~\cite{SpinGlassRev}, and any known algorithm for finding its ground state is ineffective for moderate network sizes.

At finite temperatures, the famous Parisi's replica symmetry-breaking solution of the SKSG model~\cite{Parisi_1980} reveals that spin systems can exist in two phases: spin-glass at low temperature, and paramagnetic at high temperature.
Sampling in the paramagnetic phase is easy, as expectation values are dominated by thermal noise.
However, this also means that a BM operating in such phase is unable to faithfully reproduce any probability distribution different than the aforementioned thermal noise.
On the contrary, sampling in the spin-glass phase is difficult as the free energy landscape is composed of local minima separated by large energy barriers.
Moreover, as the temperature is lowered, more minima and barriers arise.
Eventually at zero temperature their number scales exponentially with the size of the system, giving rise to an ultrametric landscape~\cite{SGultrametricity,ultrametricity}.
In this landscape, simple MCMC sampling algorithms which imitate thermal fluctuations, like Gibbs sampling, get trapped in the phase space (i.e. they present poor mixing) due to the height of the free energy barriers to be overcome.
On the other hand, global algorithms have to deal with an exponential number of local minima, leading to exponentially large times for reaching the solution.
Note that this is not a deficiency of particular sampling algorithms, but rather a manifestation of the glassy nature of the spin system.
Indeed, as the temperature approaches zero, sampling must be more and more difficult since finding the ground state of a spin glass at zero temperature is an NP-complete problem.

The standard way of avoiding spin-glass complexity in BMs consists in reducing the magnitude of the initial weights~\cite{HintonRBMguidelines} such that the effects of temperature will dominate and the system will be in a paramagnetic phase.
As a trade-off, the training signal is weaker as it is masked by thermal noise.
This can be especially troublesome in deeper layers of, e.g., deep BMs.
Indeed, the efficient training of deep BMs is perhaps the biggest challenge in the area of energy-based models.

Recent advances in analog quantum computers have led to another way of dealing with spin-glass complexity, namely quantum-assisted sampling~\cite{2018Perdomo_OrtizQST,2018AminPRX}.
The use of quantum resources for sampling BMs is advocated by theorems stating the intractability of sampling in BMs~\cite{long2010intractable}, which go beyond the case where the associated Ising system is in a spin-glass phase.

Given the above, we take a different approach: instead of dealing with intractable models---inside or outside a spin-glass phase---we define regularized models where low-energy states are readily accessible.
It is important to point out that, for any given probability distribution, there is a large number of different BMs which can approximate it~\cite{younes1996synchronous}.
We argue, and support experimentally in Section~\ref{sec:results}, that the models with regularized weights arising from RA is within such set and hence one can avoid dealing with intractable ones without losses in representability power.

\subsection{The initialization of BMs as an SKSG model is a bottleneck of training}
The paramount difficulty of sampling a spin-glass at low temperatures, and the thermal noise that arises when one attempts to solve that problem by moving to the paramagnetic phase, beg the question: is there a strong reason why one would need to initialize BMs with weights leading to an SKSG model in the first place?
Below we answer this question in the negative.

Since BMs in the paramagnetic phase cannot faithfully represent any probability distribution but those close to thermal noise, let us focus our discussion on the SKSG phase.
Indeed, the key point we raise is that BMs reproducing typical training data probability distributions are associated to Ising models outside the SKSG phase.
After a theoretical ``perfect'' training of a BM (note that this is not desired in practical scenarios because it corresponds to the memorization of the training set), the only lowest-energy neuron configurations should be those corresponding to training datapoints, all other having significant higher energies. From the theory of Hopfield networks~\cite{Cover1965,Gardner_1988} it is known that the maximum number of memorized and \textit{retrievable} configurations scales linearly with the number of neurons in the system, and not exponentially, as in the case of SKSGs. The SKSG phase has simply too many minima to allow for stable memorization.
This argument carries over to the case of practical scenarios, where the main objective is generalization instead of memorization, and successful training means that the neuron configurations representing retrievable \textit{features} of many training datapoints conform the low-energy spectrum of the associated Ising model.

Furthermore, one can analyze the spin-glass behavior of a spin system by studying whether the distribution of samples drawn from it is ultrametric~\cite{ultrametricity}.
If the training data are not strongly ultrametrically distributed (which is the case for standard datasets), the distribution of sampled outputs of a BM properly trained on it should neither be ultrametric.
On the contrary, BMs in the SKSG phase necessarily produce strongly ultrametrically distributed outputs, much more than the training data~\cite{BMultrametricity}. In order to reduce the ultrametricity of the outputs, BMs initialized in an SKSG phase must abandon it throughout training.

These arguments strongly suggest that, even if one initializes a BM as the SKSG model, the training process will drive the weights outside it, and thus, the glassy model is an unnecessary feature of current initialization and training methods.
The experiments we report on in Section~\ref{sec:results} support such scenario: during training of standard RBM models, the ability to access the low-energy states via Gibbs sampling rises dramatically in the later phases of training, while the spectral decomposition of the model weights, detailed in Section~\ref{sec:results:SVD}, shows a departure from the SKSG model.

\subsection{The rationale behind RA}
\label{sec:rationale_reg}
The SKSG phase is related to the so-called phenomenon of \textit{spin frustration}, which occurs when there is no configuration that minimizes the energy of all pairwise interactions at the same time.
The difficulty of finding the ground state and the exponential number of low-energy minima characteristic of the SKSG model are directly related to a strong frustration, which typically appears when the couplings between the neurons in the model are randomly distributed.

However, not all models with random weights exhibit frustration and spin-glass phenomenology.
In Ref.~\cite{MattisFrustration}, Mattis introduced a model with random weights but no frustration: he considered a set of $N$ variables $\xi_j$ taking values $\pm 1$, and defined the interaction between spins as $W_{ij}=\xi_i\xi_j$.
Importantly, the configurations $\bm \xi=(\xi_1,\dots,\xi_N)$ and $-\bm\xi$ correspond to the unique ground states of the spin system with couplings given by $W_{ij}$, as all pairwise interaction energies are minimized.
Furthermore, sampling in such model is easy.
RA, as given by Eq.~\eqref{eq:hebbian_rule}, can be seen as a generalization of Mattis' approach.
Indeed, it interpolates between Mattis' original procedure, where for $K\,{=}\,1,2$ the system is unfrustrated but with poor plasticity (so it cannot learn complex datasets), and $K\,{\rightarrow}\,\infty$ where the weights are uncorrelated random Gaussian variables leading to the SKSG model where standard RBMs typically begin training.

The properties of Ising models with random RA couplings have been studied in the context of the Hopfield model of associative memory~\cite{1974little,1982Hopfield}.
In particular, it is well known that the ratio $K/N$ is the parameter that determines the phase of the associated Ising system~\cite{1985Amit,amit_1989}.
In general, there exists a threshold value beyond which the model at low temperatures is in a spin-glass phase where computing or approximating $\langle\cdot\rangle_\textrm{model}$ is hard.
In contrast, below the threshold it is easy to access to the low-energy configurations and thus $\langle\cdot\rangle_\textrm{model}$ is easy to approximate.
This is the motivation to suggest, as a general procedure, to first choose a number of patterns $K$ large enough to faithfully learn the relevant features of the data, and after that the number $H$ that makes the ratio $K/(V\,{+}\,H)$ low enough to avoid the spin-glass phase.

\section{Experiments}\label{sec:results}
We proceed now to analyze the performance of RA-BMs and training using PID---with and without supplementary Gibbs sampling---in learning different datasets.
To compare it with BMs trained through standard methods we will focus on RBM architectures.
The models employed in this section, which can be found in Ref.~\cite{compapp}, are implemented in PyTorch~\cite{pytorch} via the \textit{ebm-torch} module~\cite{ebm-torch}, and run on  a workstation running Ubuntu Server 16.04 LTS, equipped with an Intel Xeon v3 E5-1660 (3GHz) CPU, 64GB of RAM, and an NVIDIA Titan Xp 12GB GPU card.

\subsection{Benchmark with exact training: 4x4 Bars}\label{sec:results:smallbas}
As a first example we trained RBMs with a small number of visible neurons, $V\,{=}\,16$, and relative to that, a large number of hidden neurons, $H\,{=}\,1\,000$.
The small $V$, along with the restricted architecture, allows for the exact calculation of the loss function, Eq.~\eqref{eq:nll}, and thus to employ exact stochastic gradient descent.
Furthermore, the ground state energy can be exactly determined at any moment of training, irrespective of whether the system is in an SKSG phase or not.
Therefore, we can meaningfully compare RAPID with the training of exactly solved RBMs.
Moreover, we also compare it against standard methods employed for training larger RBMs such as Contrastive Divergence (CD) and Persistent Contrastive Divergence (PCD) with 10 Gibbs steps and, in the case of PCD, 2048 fantasy particles.
Our initial benchmark problem is learning the Bars dataset, consisting of $4\,{\times}\, 4$ images with full vertical bars, containing a total of $14$ inequivalent images.
For such example, we choose $K=8$ for RAPID.

\subsubsection{Gibbs sampling ground state accessibility}
From the training perspective, the most important aspect of the model being in the SKSG regime or not is how hard it is to obtain a faithful distribution of states via sampling.
To estimate this, we assess the ease of reaching the ground state (GS) via Gibbs sampling starting from random visible configurations.
For doing so, we initialize the visible neurons in a random configuration and we use Gibbs sampling to extract a representative configuration of the model.
We perform $10$ Gibbs steps, after which we calculate the energy of the resulting configuration in the visible and hidden layers.
We define the ratio of such energy to the true GS energy as the {\it Gibbs sampling GS accessibility} or, shortly, the \textit{Gibbs accessibility}.

\begin{figure}
    \centering\includegraphics[width=1.05\columnwidth]{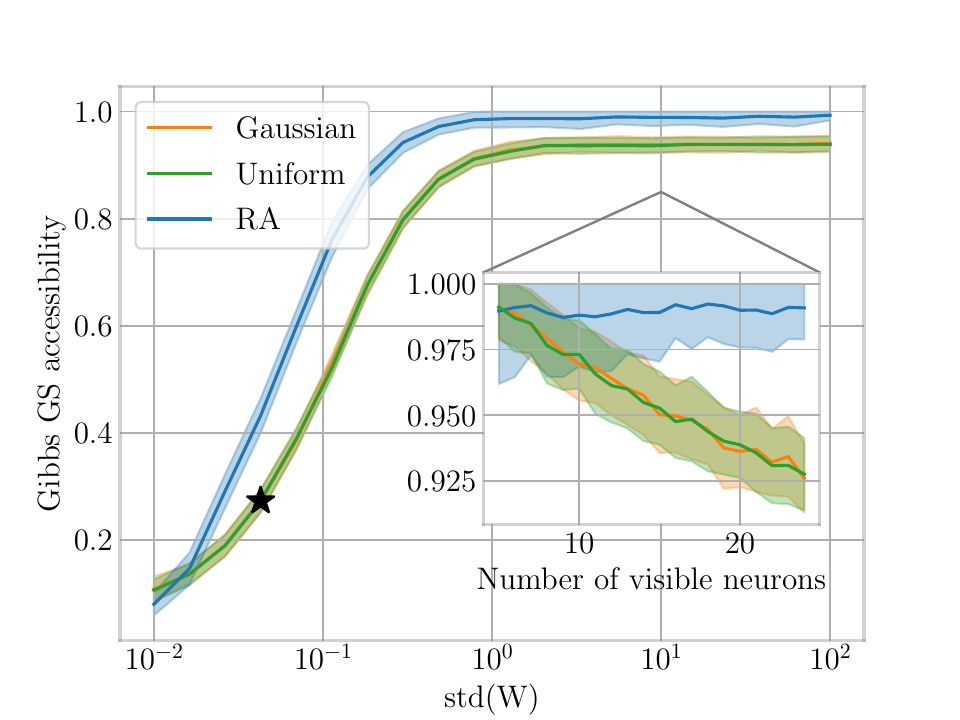}
	\caption{\textbf{Comparison of untrained models:} Gibbs accessibility for untrained models as a function of the measured standard deviation of the weights.
	The blue line represents RA-RBM models with $V=20$ and $H=1\,000$, whose weights are computed via Eq.~\eqref{eq:hebbian_rule} with $K=10$ Bernoulli-random patterns.
	The remaining lines represent standard, unrestricted RBM models of the same size, with weights sampled from (orange) a Gaussian random distribution with mean zero and (green) a uniform distribution centered at zero.
	The black star denotes the point corresponding to the Glorot initialization~\cite{glorot} for the specified values of $V$ and $H$.
	The shaded areas represent the standard deviations over 100 independent executions.
	The inset shows the Gibbs accessibility as a function of the number of visible neurons in the model, at the (fixed) standard deviation of weights equal to $10$.
	}
	\label{fig:gibbs_access_notrain}
\end{figure}

In Fig.~\ref{fig:gibbs_access_notrain} we show the Gibbs accessibility for untrained, randomly initialized models with varying $V$ and constant $H$, as a function of the standard deviation of the models' weights.
The standard deviation of weights defines the scale of system energies, and through it, the impact of temperature and thermal fluctuations on the model's dynamics.
Note that, from Eq.~\eqref{eq:boltz_prob}, the temperature of the associated Boltzmann distribution is implicitly set to $1$.
Thus, from now on, any mention to high and low temperature will be referring to low and high standard deviation of the weights' distribution, respectively.

The first notable observation is that, except for extremely small energy scales corresponding to models in paramagnetic phases due to the very high temperatures, the Gibbs accessibility is higher for RA-RBMs than for RBMs initialized in a standard way.
Therefore, sampling low-energy configurations from models with RA is easier than sampling low-energy configurations from unregularized models.
Moreover, the difficulty of sampling low-energy configurations in unregularized models is not affected by whether the values of the weights are drawn from Gaussian or uniform distributions.

Next, focusing on the variation of the Gibbs accessibility with the energy scale, one observes that small weights lead to system dynamics dominated by thermal fluctuations, and thus exploring high-energy configurations.
In such regime, all models are in their respective paramagnetic phases and the Gibbs accessibility is low for all of them.
Learning in a regime of small weights is usually slow, but typical guidelines for training RBMs suggest to start in this regime~\cite{HintonRBMguidelines}.
Even modern approaches to weight initialization suggest initial parameter values that give rise to models where the access to the low-energy spectrum via sampling is poor.
For example, the well-known Glorot initialization~\cite{glorot} (denoted by a star in Fig.~\ref{fig:gibbs_access_notrain}) generates initial models where Gibbs sampling is only capable of reaching configurations whose energy is not lower than three times the energy of the ground state in small models.

\begin{figure*}[ht!]
    \centering
    \subfloat[\label{fig:gibbs_access}]{\includegraphics[width=0.43\textwidth]{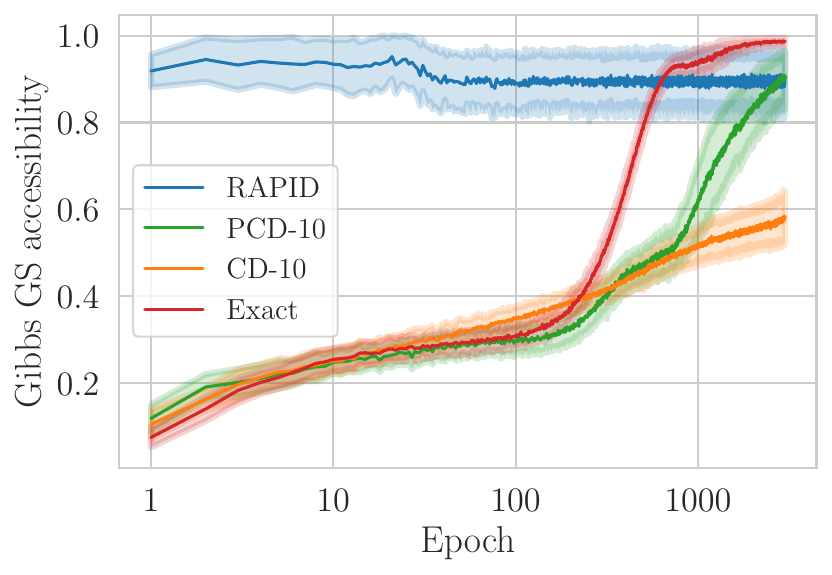}}
    \hskip 0.5in
    \subfloat[\label{fig:pattern_access}]{\includegraphics[width=0.43\textwidth]{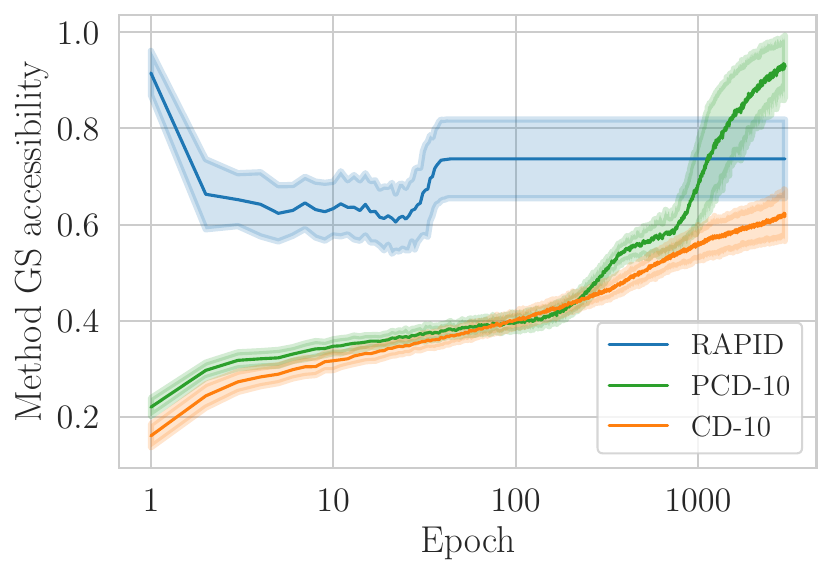}}
	\caption{\textbf{Characterization of the low-energy space:} Accessibility of the GS in RBMs, using different training methods. (a) Smallest energy, relative to the ground state energy, of the configuration obtained after $10$ steps of Gibbs sampling, beginning from random visible configurations. (b) The method accessibility measures how well the negative phase captures the low-energy behavior of the exact Boltzmann averages, by comparing the energy of the lowest-energy configuration employed to compute the negative phase with that of the ground state. In all cases, the models tested have $V\,{=}\,16$ visible and $H\,{=}\,1\,000$ hidden neurons, and are trained in the $4\,{\times}\,4$ Bars dataset. For the case of RAPID (in blue), we employ $K\,{=}\,8$ patterns. The shaded regions denote the standard deviations after $100$ instances of independent training.}
	\label{fig:sampling}
\end{figure*}

As the standard deviation of weights is increased, the impact of thermal fluctuations decreases.
Eventually, for large weights the thermal fluctuations are negligible.
In this regime, the Gibbs accessibility is independent of the energy scale, as shown by the plateau in Fig.~\ref{fig:gibbs_access_notrain}.
While the behavior of standard RBM and RA-RBM models is similar when increasing the energy scale for a fixed number of neurons, the fact that the value at the plateau is different is a signature of the fact that standard RBM models are in an SKSG phase where reaching the ground state through sampling is more difficult than in RA-RBM models, which are instead in a ``few-minima'' phase where the number of energy minima scales polynomially with the system size, instead of exponentially.
Crucially, this different behaviour accentuates when, for energy scales in the plateau, one considers models with an increasing number of neurons (shown in the inset of Fig.~\ref{fig:gibbs_access_notrain}).
In the case of standard RBMs, the system is in a low-temperature SKSG phase where sampling the ground state is hard due to the existence of an exponential number of minima.
Indeed, the Gibbs accessibility quickly decreases when one increases $V$, as a consequence of the problem of finding the ground state in an SKSG phase being NP-complete.
Contrary to that, for our regularized RA-RBM the Gibbs accessibility stays constant when increasing $V$.
This strongly suggests that such model is not in the SKSG phase, but in a regime at low temperature where sampling low-energy configurations is easy while the signal is not dampened with thermal fluctuations, and where the number of minima is controlled not by $H$ or $V$, but by $K$.

Next, we analyze how the Gibbs accessibility varies with training, which is depicted in Fig.~\ref{fig:gibbs_access}.
For RBMs trained with CD, PCD, and exact gradients, the models are initialized in accordance to the standard procedure~\cite{HintonRBMguidelines}, thus being initially in a paramagnetic phase at high temperature.
As discussed above, this initialization has the consequence that, during the first epochs of training, Gibbs sampling does not reach low-energy configurations.
This effect is prominent in Fig.~\ref{fig:gibbs_access}, and in stark contrast to the case of RA-RBMs, for which Eq.~\eqref{eq:hebbian_rule} initializes the model in a phase where the ground state is easily accessible via Gibbs sampling.

After training, Fig.~\ref{fig:gibbs_access} shows that all standard RBM models end up in a regime where Gibbs sampling is efficiently reaching the low-energy sector.
The speed at which they reach this regime is directly related to the quality of the estimation of the negative phase, this is, to the ability of drawing samples according to the Boltzmann distribution of Eq.~\eqref{eq:boltz_prob}.
In contrast, RA-RBM models are always in a regime of good sampling, which allows for large reductions in the number of epochs needed for successful training (see Sec.~\ref{sec:results:hd}).

\subsubsection{Method ground state accessibility}
In Fig.~\ref{fig:pattern_access} we consider a quantity more relevant during the training process: the proximity of the configurations employed by each method to compute the negative phase, $\left<\partial_\theta E\right>_\mathrm{model}$, to the respective ground states.
For the various training methods, we define the \textit{method GS accessibility} as the ratio of the lowest-energy configuration employed in the computation of the negative phase to the ground state energy.
Note that, when employing the exact gradients, we have an explicit expression for $P_\mathrm{model}$ and therefore there is no need of taking any samples from the model.
This is the reason why there is no curve in Fig.~\ref{fig:pattern_access} for the exact training method.

\begin{figure*}[ht!]
    \centering
	\subfloat[\label{fig:hd_train}]{\includegraphics[width=0.43\textwidth]{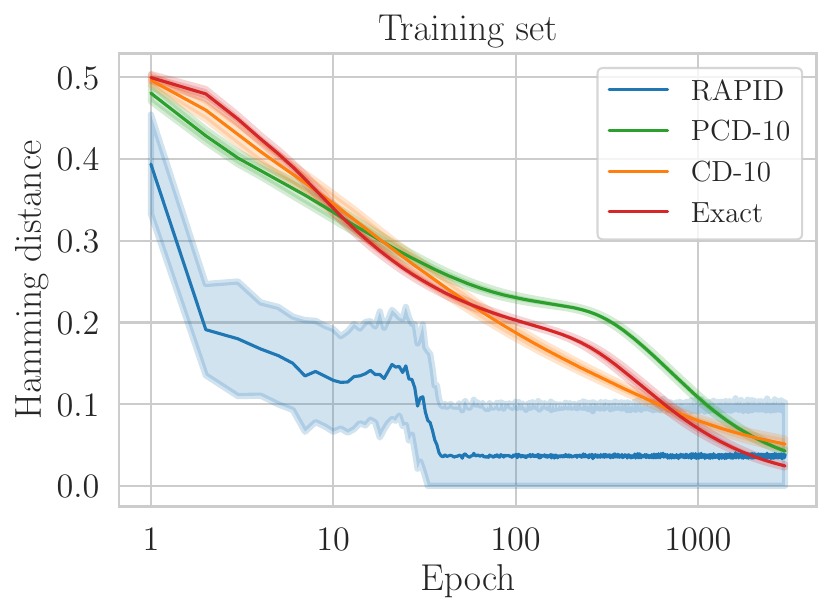}}
    \hskip 0.5in
    \subfloat[\label{fig:hd_test}]{\includegraphics[width=0.43\textwidth]{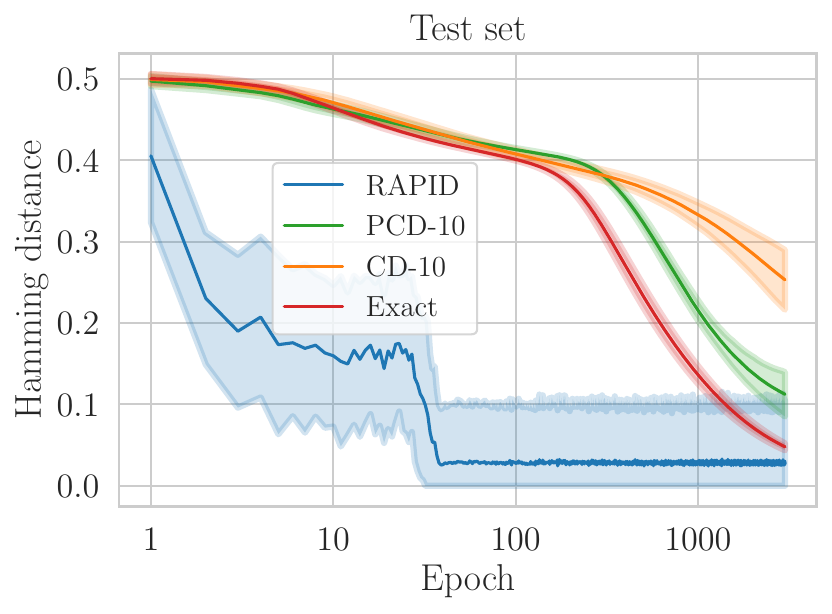}}
    \caption{\textbf{Learning accuracy:} Hamming distance between reconstructions of partial images and expected results in the (a) training and (b) test sets of the $4\,{\times}\,4$ Bars dataset. The shaded areas around the lines denote the standard deviation of $100$ independent training instances. The parameters of the models are the same as those in Fig.~\ref{fig:sampling} ($V\,{=}\,16$, $H\,{=}\,1\,000$, and $K\,{=}\,8$ in the case of RA).
    }\label{fig:learning}
\end{figure*}

For CD, the Gibbs sampling and method accessibilities are very similar, since the method for computing both is, in essence, the same.
The method accessibility of Fig.~\ref{fig:pattern_access} is slightly better due to the fact that, in that case, the initial configurations before sampling are images from the training set instead of the random configurations used when computing the Gibbs accessibility of Fig.~\ref{fig:gibbs_access}.
A similar phenomenon can be observed in the curves for PCD.
In this case, the method accessibility is better than the Gibbs accessibility due to the fact that the fantasy particles employed in the sampling are always close to the ground state.
In the case of RAPID, it is apparent that, at late stages of training, conventional methods seem to provide a better characterization of the ground space than the pure PID defined in Eq.~\eqref{eq:pid}.
Nevertheless, this is counteracted by the greatly better characterization provided by PID in the initial training epochs.
Indeed, this improved accessibility to the low-energy space of configurations at the initial stages of training leads, as explicitly shown in Fig.~\ref{fig:learning}, to achieve successful learning much before the conventional methods surpass PID in method accessibility.

We observe that for PID the method accessibility does not improve with training.
While, as we show below, this is not an issue for small datasets, it may constitute a problem when scaling the method and using it for learning more complex data.
We note that the results of Fig.~\ref{fig:sampling} are obtained with the pure PID described in Eq.~\eqref{eq:pid}, where no MCMC is employed for computing $\langle\cdot\rangle_\textrm{model}$.
A straightforward way of improving the method accessibility is thus employing the patterns $\{\bm\xi^{(k)}\}_k$ as seeds for MCMC methods.
In Section~\ref{sec:results:complex} we employ this combination of PID and Gibbs sampling when learning the MNIST dataset.

\subsubsection{Learning and generalization accuracy}\label{sec:results:hd}
In order to quantify the performance on learning the Bars dataset, we ask the models to reconstruct corrupted images (see Appendix~\ref{app:task} for the details of this task).
Following the standard procedure of unsupervised learning, we divide the dataset into two sets: a training set consisting of 10 images, and a test set containing the remaining 4 images.
In the case of spin values and no local neuron biases, the energies of configurations $\bm v$ and $-\bm v$ are the same.
Therefore, in order to ensure that there is no information leakage from the training set to the test set, we design them in such a way that the negative of every configuration in the training set is also in the training set, and the negative of every configuration in the test set is also in the test set.

In Figs.~\ref{fig:hd_train} and~\ref{fig:hd_test} we depict how the reconstruction of the training and test sets, respectively, evolve during training.
For the case of simple datasets such as those employed, the Hamming distance (HD) provides a very good assessment of the quality of training, despite of it not being the quantity being optimized [which, recall, is given by Eq.~\eqref{eq:nll}].
One observes that: (i)~the different training methods for standard RBMs lead to very similar memorization (the reduction of the HD in the training set) while generalization (the reduction of the HD in the test set) is faster with improved approximations of the negative phase, and
(ii)~there is almost no difference between the performance on memorization and generalization when employing RAPID.
This is a clear indication that the corresponding model not only truly learns, but also it does so very efficiently.

One should deal with these results with care, as they need not imply that RAPID allows for faster learning---in terms of the number of epochs---when compared to methods of training standard models (nevertheless, as we show in Appendix~\ref{app:cost}, PID presents a speedup in the complexity of the computation of each update).
We showed in Fig.~\ref{fig:gibbs_access_notrain} that it is possible to initialize standard models outside the SKSG regime, for instance by increasing the scale of the weights, making the starting points in Fig.~\ref{fig:sampling} much closer to the initial point of RA.
The implications of such procedure, and the impact of large weights and of $H\gg V$ (which is typically necessary for having $K\gg N$) in other quantities useful for tracking training---in both RA and standard models---are not yet fully understood but are important aspects that will provide a better assessment of the performance of energy-based unsupervised learning methods~\cite{further}.

\subsubsection{Spectral decomposition of trained models}\label{sec:results:SVD}
\begin{figure}[h]
    \centering\includegraphics[width=0.9\columnwidth]{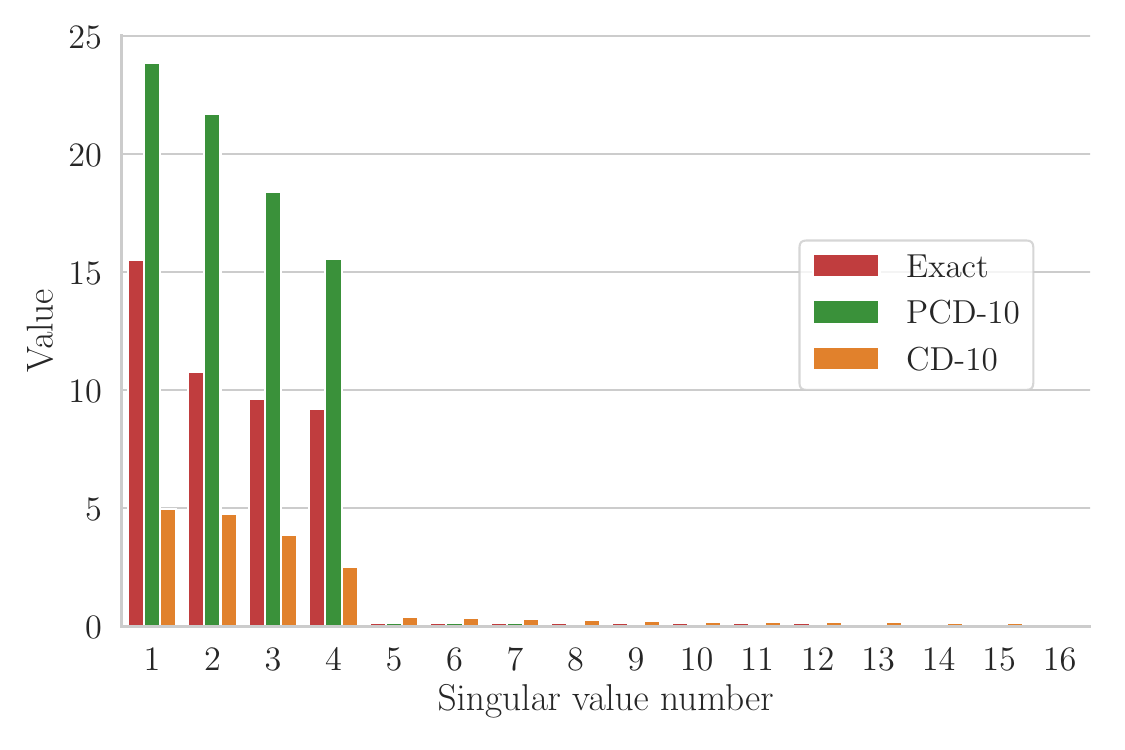}
	\caption{\textbf{Spectral decomposition of the weight matrices:} SVD of the weights after training RBM architectures after learning the $4\times 4$ Bars dataset, using different training methods. Note that, regardless of the training method, if the RBM learns the dataset then there is a small number of relevant singular values.}
	\label{fig:singular}
\end{figure}

Regardless of the above, one may still wonder how similar are the RA- and standard RBM models after training.
For doing so, we perform a singular value decomposition (SVD) of the weight matrices of trained models.
Such results are shown in Fig.~\ref{fig:singular}.
Clearly, in all cases, four large singular values stand out.
Interestingly, the form of the SVD of the weight matrix (see, for instance, \cite[Eq.~(10)]{SVDRBM}) invokes Eq.~\eqref{eq:hebbian_rule}, such that a clear analogy between the patterns employed in RA and the SVD eigenvectors can be drawn.
The four large singular values observed in Fig.~\ref{fig:singular} suggest that only $\approx 4$ patterns should be sufficient to describe the weight matrix of a standard RBM trained in the $4\,{\times}\,4$ Bars dataset, in all cases of training methods studied.
We note that an SVD analysis of standard RBMs trained on MNIST has been already performed in Ref.~\cite{SVDRBM}, where it was reported that the SVD spectrum develops a tail of relatively few but large singular values.
Taking also into account that the standard RBMs presented in Fig.~\ref{fig:gibbs_access} evolved towards a regime of easy sampling, we can interpret that the training of BMs drives the weights to a low-temperature but non-SKSG phase, and thus, that \textit{initializing Boltzmann machines as an SKSG model is an unnecessary and avoidable bottleneck}.
These results also show that the RA-RBM may be regarded as an actual general model of trained standard RBMs.

\subsection{Increasing complexity: 12x12 BAS and MNIST}\label{sec:results:complex}
\begin{figure}[t]
    \centering
    \begin{overpic}[width=0.41\textwidth]{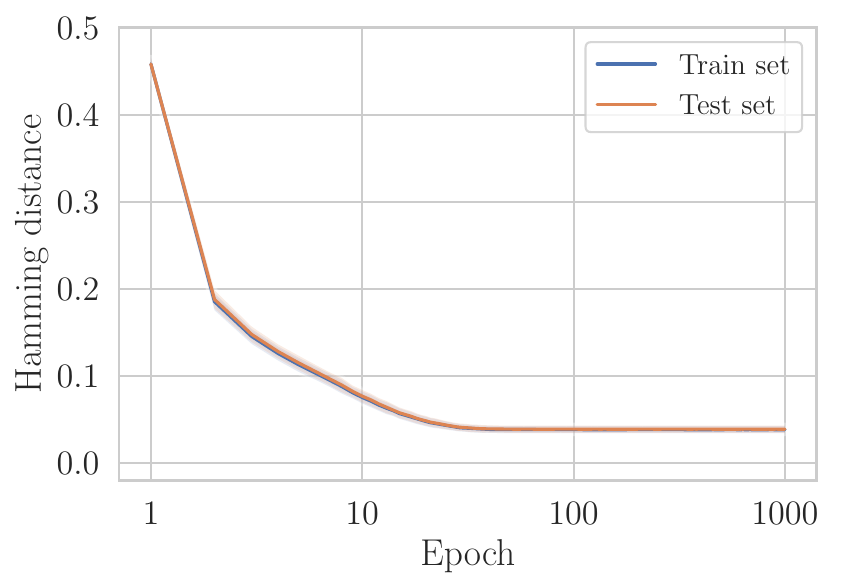}
        \put(49,23){\includegraphics[scale=0.12]{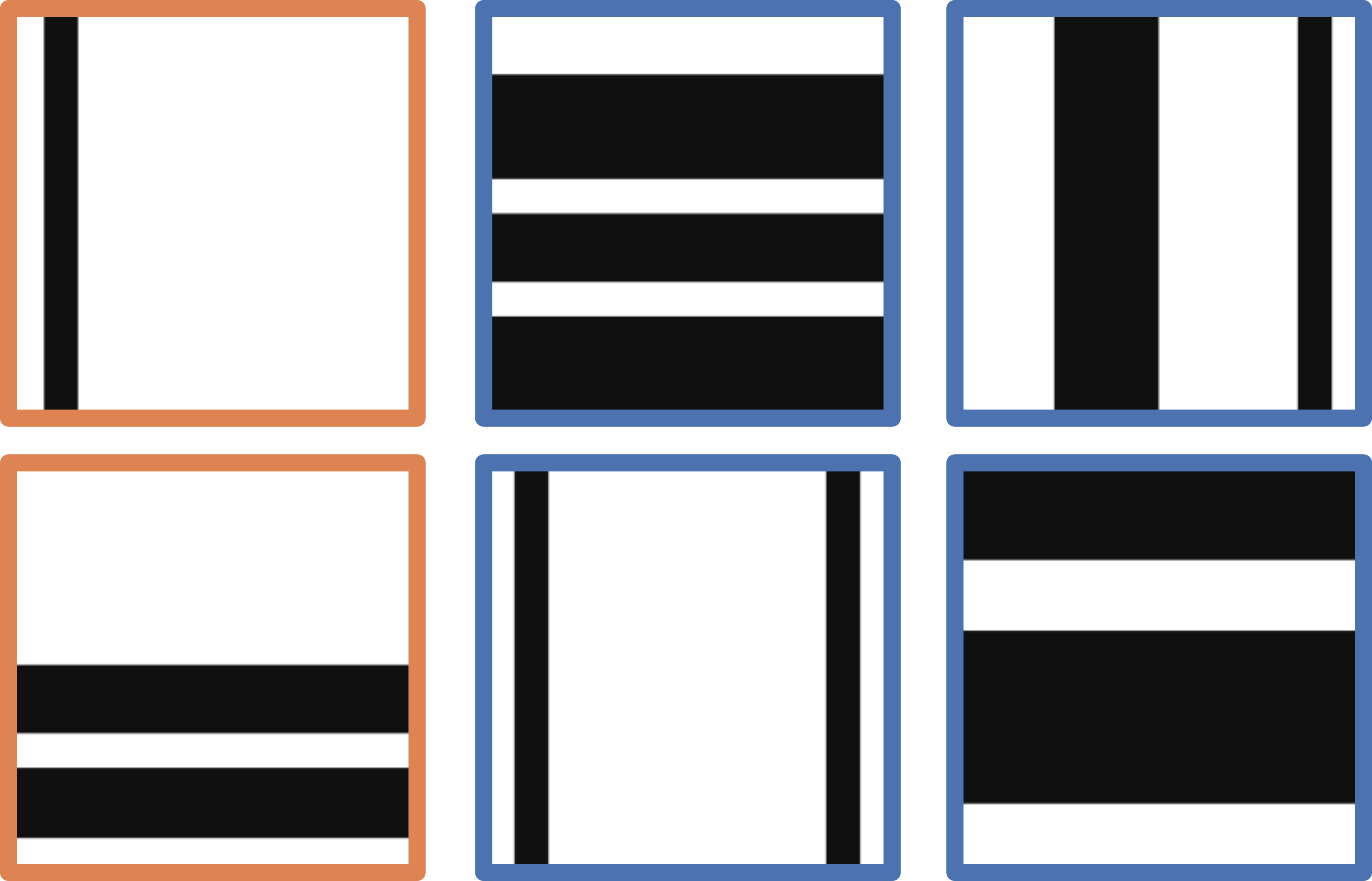}}
    \end{overpic}
\caption{\textbf{RAPID in 12x12 BAS:} Hamming distance between reconstructed images and expected results for the $12\,{\times}\,12$ Bars and Stripes dataset. The models employed have $V\,{=}\,144$, $H\,{=}\,1\,000$, $K\,{=}\,40$, and a batch size of 120, and was trained with 80\% of the dataset ($6\,550$ images). The shaded regions denote the standard deviations in 100 independent training instances. The inset shows instances sampled from the model. The leftmost samples, surrounded in orange, were not part of the training set. }
	\label{fig:hd_bas_12x12}
\end{figure}

We now proceed to apply RAPID to the unsupervised learning of more complex datasets.
First, we consider the $12\,{\times}12$-pixel Bars and Stripes (BAS) dataset, which consists of $8\,188$ images containing only vertical bars or horizontal stripes, separated in a training and test set with 80/20 ratio.
As the complexity of the problem to solve increases, one needs to increase the number of auxiliary patterns $K$ and, if necessary, the number of hidden neurons $H$.
In Fig.~\ref{fig:hd_bas_12x12} we show the results for the HD of reconstructed images.
The HD for the training and test sets decrease parallel to each other, proving that the model trained is not just memorizing the images of the training set, but learning their fundamental features and being able to generalize the results to the test set.
The inset shows images generated from sampling the model starting from a random initial visible configuration. The border color indicates whether the generated image is contained in the train or test set.
As a powerful print of the generalization power of the model, we see that not only it reconstructs satisfactorily corrupted unknown images (the results on the HD), but moreover it is able to generate images that were not contained in the training set.

\begin{figure*}[ht!]
    \centering
    \subfloat[\label{fig:mnist_nll}]{\includegraphics[width=0.43\textwidth]{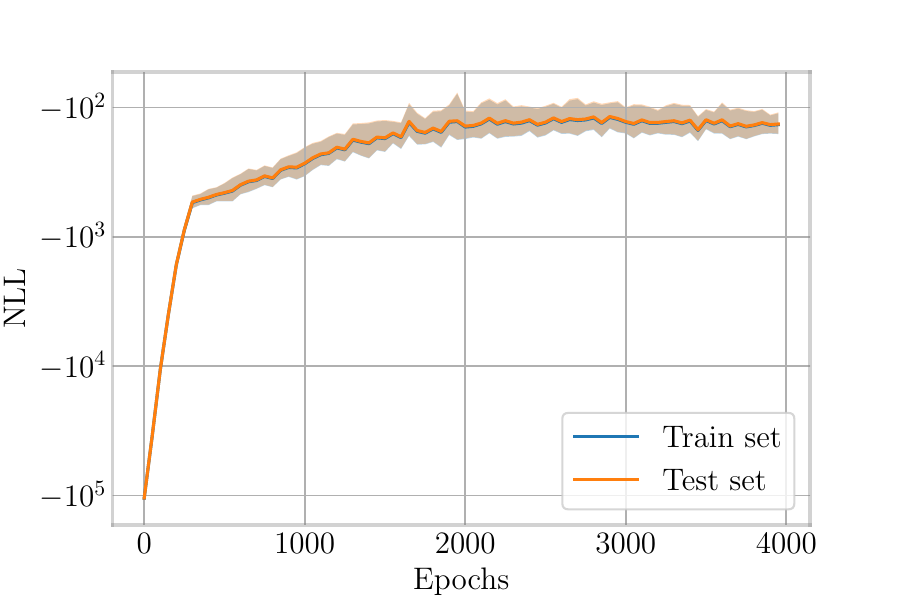}}
    \hskip 0.3in
    \subfloat[\label{fig:mnist_nll_zoom}]{\includegraphics[width=0.43\textwidth]{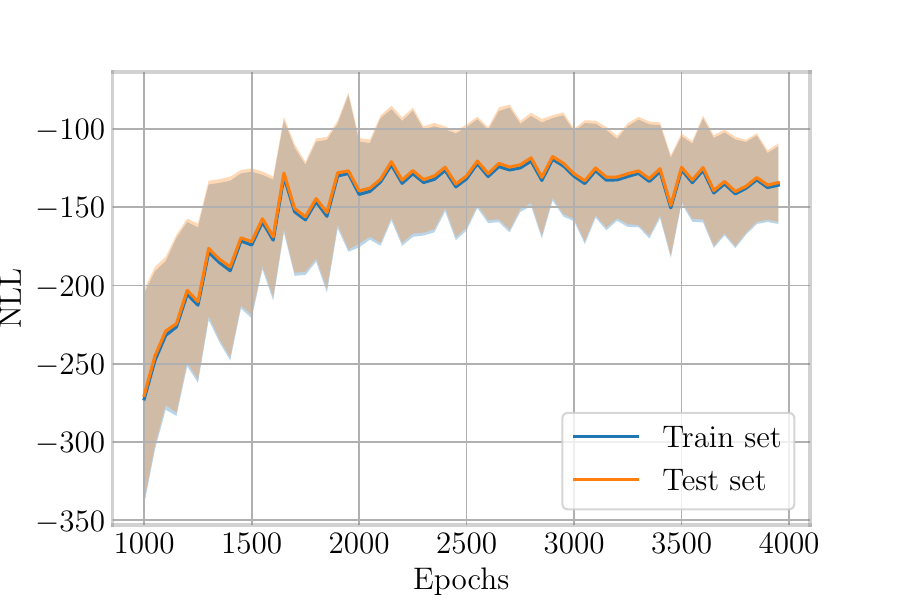}}
    \hskip 0.3in
    \subfloat[\label{fig:mnist_samples}]{\includegraphics[width=0.3\textwidth]{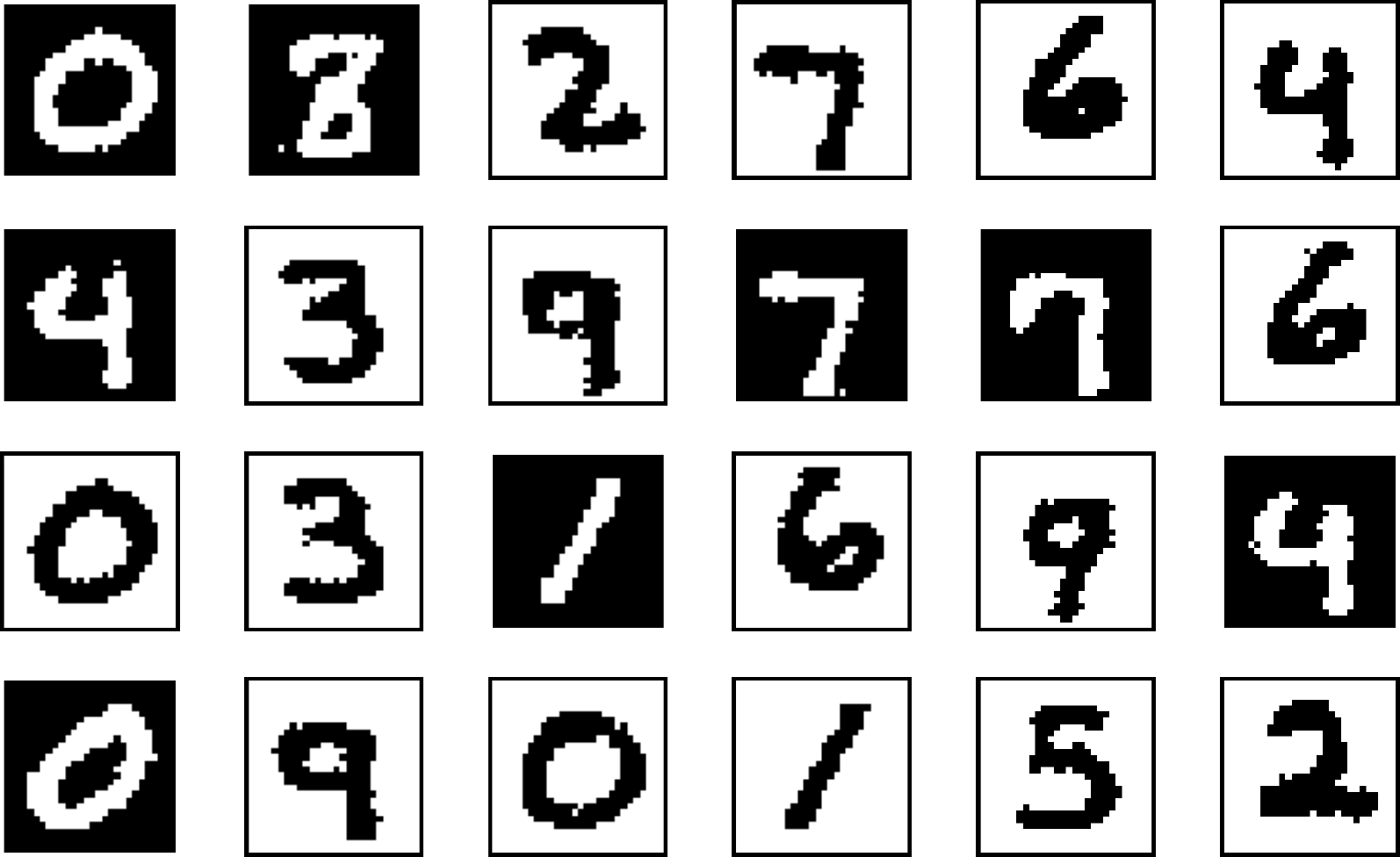}}

    \caption{\textbf{RAPID in MNIST:} (a) Negative log-likelihood, estimated via Annealed Importance Sampling~\cite{rais}, for the train and test sets. Since AIS provides a lower bound to the partition function, the curves depicted are lower bounds to the true NLL. (b) Zoom-in on the region of late training. The parameters used for both (a) and (b) are $H\,{=}\,3\,000$, $K\,{=}\,190$, $LR\,{=}\,0.015$ and $10$ Gibbs steps to the patterns, as explained in Sec.~\ref{sec:results:complex}. The shaded areas around the lines denote the standard deviation of 25 independent training instances. (c) Instances sampled from a trained model. The parameters of the latter are $H\,{=}\,3\,000$, $K\,{=}\,290$, $LR\,{=}\,0.015$ and 10 Gibbs steps to the patterns.}
    \label{fig:mnist}
\end{figure*}

As a final example, we employ RAPID to train an RBM in a binarized version of the MNIST dataset.
Here, the complexity of the dataset is much higher than in the BAS example, and so it requires to increase the size of the model both in terms of $K$ and $H$, which nevertheless does not cause an important impact in terms of computation speed.
This is the first case in which we observe that the low-energy space characterization of PID is not satisfactory to perform proper learning (recall Fig.~\ref{fig:pattern_access}).
In fact, employing pure PID for computing the negative phase led to a strong overfitting.
In order to avoid this and to achieve a good approximation of $\left\langle\cdot\right\rangle_\mathrm{model}$, we employ the patterns $\{\bm \xi^{(k)}\}_k$ as initial seeds for a series of steps of Gibbs sampling.
The resulting configurations after the sampling are those employed for approximating the average under the model distribution.
This, in addition to enhancing the proximity of the patterns to the low-energy sector of the model as discussed in Section~\ref{sec:method:pid}, introduces fluctuations which are known to help to overcome overfitting.

Another important change with respect the previous implementations is that we now allow the patterns $\{\bm \xi^{(k)}\}_k$ to have continuous values, $\xi_i^{(k)} \in [-x,x]$, as we describe in Appendix~\ref{app:discrete}.
This only results in a higher plasticity of the model (i.e. a larger set of possible weights $W$) and does not imply substantial changes in the main features of RA and PID discussed previously.
Indeed, Mattis' argument that weights constructed from a single pattern give rise to a spin model without frustration applies irrespectively of the pattern taking binary or continuous values.
Therefore, RA in the case where patterns take continuous values still can be seen as a regularization that interpolates between a Mattis-like unfrustrated model for a single pattern and a frustrated spin-glass model for infinitely many patterns.
Such reduced frustration again suggests existence of only a few low-energy minima.
Indeed, in the case of a single continuous-valued pattern $\bm{\xi}$ the associated spin model has just two equivalent ground states, which correspond to $\text{sign}(\bm\xi)$ and $-\text{sign}(\bm\xi)$.
As for PID, the continuous-valued patterns cannot be understood as real spin configurations, and thus cannot be used directly in Eq.~\eqref{eq:pid}.
Yet, the reasoning above suggests that a simple binarization of the patterns produces configurations that lie in the low-energy sector.
Due to the aforementioned issues with overfitting, we instead decide to employ a more elaborated approach, using the binarization of the continuous-valued patterns as seeds of Gibbs iteration chains, whose final states correspond to the spin configurations that are used to compute the negative phase of the gradient update.

We now discuss the results of training a large RA-RBM on the MNIST dataset. On one hand, we show in Fig.~\ref{fig:mnist} how the quantity being optimized, the NLL in Eq.~\eqref{eq:nll}, evolves as training progresses. Due to the practical impossibility of exactly calculating the partition function for large models, we approximate it via Annealed Importance Sampling (AIS)~\cite{rais}. In Figs.~\ref{fig:mnist_nll} and \ref{fig:mnist_nll_zoom} we show the training progress for a model with $H=3\,000$, $K=190$, and using the patterns as initial seeds of $k=10$ Gibbs steps of CD as commented above. We see that the model is able to optimize correctly the NLL, attaining final values four orders of magnitude smaller than the initial NLL. Moreover, the NLL for the training and test datasets move parallel, showcasing the absence of overfitting in our system. We note here that the initial value for the NLL in RA-RBM models is much smaller than the average initial values in unregularized RBMs. The full extent of such initial discrepancy falls out of the scope of the current work and is left for a subsequent analysis~\cite{further}.  Nevertheless, this does not seem to affect the training of the model, and the Gibbs-assisted PID is able to correctly optimize the NLL to values comparable to, although not necessarily better than, those obtained when training standard RBMs and other generative models on the same dataset (see, for instance, Table 3 in Ref.~\cite{Uria2016Neural}). In this sense, it must be stressed that the experiments shown do not have as primary goal to outperform the current state of the art, but rather to illustrate the capabilities of RA models.

To further showcase the fact that RA-RBM models can learn complex datasets such as the MNIST dataset, we show in Fig.~\ref{fig:mnist_samples} a number of samples generated via Gibbs sampling from an RA-RBM model with the same parameters as above but with $K=290$. While we have consistently observed that lower NLL values are achieved by models with lower $K$, we also observe that, in terms of visual quality, models with larger values of $K$ generate better samples. This observation is in line with previous analyses on performance metrics of generative models: while both characteristics (the optimization of the NLL and visually accurate samples) are indicators of learning, they do not necessarily correlate with each other~\cite{theis2016evaluation}. The fact that both indicators (the trained models generating good-quality images, and the loss being minimized in the training set but more importantly in the test set) are present in trained RA-RBMs is a powerful sign that models with RA are capable of learning complex datasets.

Moreover, we provide below a third indication of learning.
For each visible configuration corresponding to an image of the binarized MNIST training set, we produce a sample of the configuration of the hidden layer in a trained RA-RBM according the corresponding conditional distribution, and perform logistic regression on the dataset so obtained to the corresponding labels.
By repeating this procedure for 100 independent RA-RBMs trained with the same hyperparameters, the corresponding logistic regression classifiers have an average accuracy of $\mathbf{95.47\pm0.19\%}$ in the training set and of $\mathbf{93.84\pm0.19\%}$ in the test set, demonstrating that the hidden layers of the RA-RBMs have learnt to distill the relevant information of the samples of the dataset.
As a comparison, carrying out the same procedure with standard RBMs~\cite{ebm-torch} with $300$ hidden neurons produces classifiers with $93.01\pm0.10\%$ average accuracy on the training set and $92.04\pm0.19\%$ average accuracy on the test set.

\subsection{Analysis of the trained patterns}
\label{sec:pattern:analysis}
Weights built from patterns like in Eq.~\eqref{eq:hebbian_rule} have been extensively studied in the context of the Hopfield model. Here, for the first time, we use a similar construction in BMs. Since both models are closely related it is natural to perceive the patterns used in RA as analogs of the patterns of the Hopfield model. However, they work quite differently, for the two reasons we explain below.

The first difference arises when analyzing the goal of both models.
While BMs aim at generating new samples as similar as possible to the those in the training dataset, the aim of the standard Hopfield model is to memorize as many samples from such dataset.
In the latter, the figure of merit is the number of \textit{retrievable} samples.
To achieve such goal, the samples of the dataset can be used as patterns from which the weights are generated, hence requiring one pattern per image in the training set in order to store the full dataset.
The total number of memorized data is then directly related to the number of patterns and neurons of the model, and they are retrieved via the network dynamics.
In the case of RA-BMs, patterns are used as a means of regularization of the weights in order to avoid the spin-glass complexity.
Therefore, the number of patterns controls the number of low-energy minima, and these minima need not correspond to single data instances. Contrarily, since the aim of BMs is to learn and generalize rather than memorize, the number of low-energy minima is usually much lower than the number of training samples.
The fact that the number of patterns is much smaller than the number of training samples is certainly the case in the experiments showcased: the models employed to learn the 12$\times$12 BAS dataset ($6\,550$ images in the training set) contain 40 patterns, and the models employed to learn the MNIST dataset ($60\,000$ images in the training set) contain 190 patterns.

Second, the crucial difference between Hopfield models and BMs is that the former have only visible units while latter have been enriched with latent hidden units.
Therefore, while patterns in the Hopfield model correspond to data instances, the patterns in our work have both visible and hidden parts, the latter lacking any correspondence to visual configurations.
The low-energy minima of RA-BMs are encoded jointly in the visible and hidden parts of the patterns used.
In this context it is interesting to note that there exist models, such as the Hybrid Boltzmann Machine of Ref.~\cite{2012Barra}, whose marginalization over hidden units is equivalent to the Hopfield model. One could thus wonder about the relations of marginalized RA-BMs with generalized Hopfield models like those in Refs.~\cite{weigt1,weigt2}. In Appendix~\ref{app:hopfield} we show that our construction, when marginalized over hidden units, is not equivalent to those cited above.

\begin{figure}
    \centering
	\subfloat[\label{fig:some_patterns}]{\includegraphics[width=0.43\textwidth]{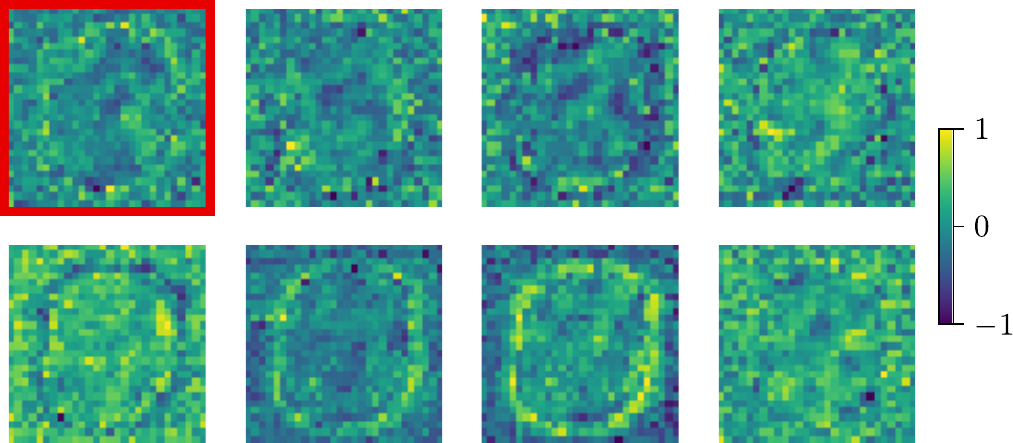}}
    \hskip 0.5in
    \subfloat[\label{fig:generation_from_patterns}]{\includegraphics[width=0.43\textwidth]{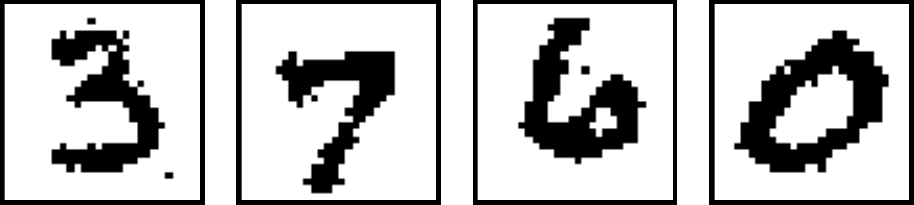}}
    \caption{\textbf{Patterns after training on MNIST:}  (a) Visible part of a random choice of the patterns of an RA-RBM after training on the MNIST dataset. They correspond to the patterns of one the models whose NLL is presented in Fig.~\ref{fig:mnist}. (b) Generated samples when the initial seed for Gibbs sampling is set to be the visible configuration highlighted in (a).
    }\label{fig:patterns}
\end{figure}

In Fig.~\ref{fig:patterns} we demonstrate the two properties described above, for an RA-RBM trained on binarized MNIST (see Sec.~\ref{sec:results:complex}). Fig.~\ref{fig:some_patterns} presents the visual parts of eight patterns randomly chosen from the $K=190$ that were used for creating the model. As it can be seen, they do not resemble any of the samples from the training dataset but rather some generalized features. It must be emphasized, as was already noted in Sec.~\ref{sec:results:complex} and in Appendix~\ref{app:discrete}, that in this case the patterns are formed of bounded continuous variables.
More importantly, as presented on Fig.~\ref{fig:generation_from_patterns}, starting Gibbs sampling from the visible part of one of the mentioned single patterns (indicated by the thick red contour) we obtain \textit{different} visible images generated by our BM, belonging to different classes of digits. Although our patterns encode low-energy minima, they do not correspond to single retrievable low-energy configurations, let alone single data instances.

\section{Discussion and remarks}\label{sec:conclusions}
We have provided two contributions to the problem of unsupervised learning of datasets with energy-based models.
First, a conceptual finding---that we support experimentally---is that initializing the parameters of energy-based models in regimes that lead to SKSG models is unnecessary, and that avoiding SKSG phenomena is possible without starting in a paramagnetic phase at high temperature where signals are dampened by thermal noise.
In supporting the above, and as a second contribution, we have developed RAPID, a combination of model choice and training method which consists of: (i) Regularizing the Axons on the model by utilizing the Hebbian rule to construct the weights of a Boltzmann machine by means of $K$ random patterns that ensure a model sufficiently expressive, and with a number of hidden neurons such that the ratio $K/N$ is kept low enough to avoid an SKSG phase at any point of training;
and (ii) employing Pattern-InDuced correlations to approximate the negative phase in the log-likelihood gradient.
We have proven in several examples that RAPID, with and without supplementary Gibbs sampling, leads to models that learn very efficiently and successfully generalize training data.

Although the cases presented are significant examples, the question on how restrictive the RA construction is for learning general probability distributions still remains.
Based on the evolution of the Gibbs accessibility during training and the singular value decomposition of weight matrices of trained RBMs shown in Sec.~\ref{sec:results:SVD} and in Ref.~\cite{SVDRBM}, we conjecture that trained RBMs are well approximated by RA-RBMs.

Models with RA seem to have the potential for very fast learning.
In the case of simple datasets ($4\,{\times}\,4$ Bars), the amount of epochs required to train models capable of reconstructing untrained images can be orders of magnitude smaller than when using standard RBMs.
In the more complex case of binarized MNIST, the comparison of learning speed with other methods is not straightforward, since the quality of the training process can have different definitions, which are all uncorrelated with each other~\cite{theis2016evaluation}.
Nevertheless, we note that the proposed RA-RBMs are capable of generating samples of arguably better quality than standard RBM models trained for more epochs (compared to, for instance, Fig. 5b in Ref.~\cite{cote2016infiniteRBM}).
Moreover, the theoretical arguments on the avoidance of SKSG phases suggest that the benefits of RA models will be more prominent in more complex scenarios, when learning datasets of higher dimensionality and when training more complex models.

This work focused on experiments with RBMs, in order to carefully compare RA and PID with standard models and training algorithms.
However, our methods are by no means restricted to RBMs, as the principles behind RA and PID can be applied to any Boltzmann machine architecture.
In fact we expect that RAPID, or variations of it, will bring the long-sought-after efficient algorithms for training deep Boltzmann machines.
However, if RAPID failed to achieve this goal, one would be able to conclude that the SKSG phenomenology is not the actual reason for the difficulty of training of deep BMs, pointing to more intricate sampling problems which could possibly signal the necessity of quantum-assisted sampling or analog computing solutions.

\paragraph*{Acknowledgments.}
M.L. and A.A. groups acknowledge the Spanish Ministry MINECO and State Research Agency AEI (FIDEUA PID2019-106901GBI00/10.13039/501100011033, Severo Ochoa grants SEV-2015-0522 and CEX2019-000910-S, FPI), the European Social Fund, Fundacio Cellex, Fundacio Mir-Puig, Generalitat de Catalunya (AGAUR Grant No. 2017 SGR 1341 and SGR 1381, CERCA program, QuantumCAT U16-011424, co-funded by ERDF Operational Program of Catalonia 2014-2020), ERC AdG NOQIA and CERQUTE, EU FEDER, MINECO-EU QUANTERA MAQS (funded by the State Research Agency AEI PCI2019-111828-2/10.13039/501100011033), the
National Science Centre, Poland-Symfonia Grant No. 2016/20/W/ST4/00314 and the AXA Chair in Quantum Information Science.
A.P.-K. acknowledges funding from Fundaci\'o Obra Social ``la Caixa'' (LCF/BQ/ES15/10360001) and the European Union's Horizon 2020 research and innovation programme - grant agreement No 648913.
G.M.-G. acknowledges funding from  Fundaci\'o Obra Social ``la Caixa'' (LCF-ICFO grant).
M.A.G.-M. acknowledges funding from the Spanish Ministry of Education and Vocational Training (MEFP) through the Beatriz Galindo program 2018 (BEAGAL18/00203).

\bibliography{biblio}
\bibliographystyle{apsrev4-1}

\appendix
\section{Parameter update rule for RA-RBM}\label{app:updaterule}
In a model with RA, the weights are not the ultimate parameters to be fixed by training.
These are, rather, the values of the auxiliary patterns $\xi_i^{(k)}$.
In this appendix we detail the calculation of the update rule for the auxiliary patterns. We focus here in the training of an RBM, just as explained in the main text.
We start by recalling that the probability of observing a state $\bm{v}$ of the visible variables is given by
\begin{equation}
\label{prob_visible}
P_\mathrm{model}(\bm v) = \frac{\sum_{\bm{h}} e^{-E(\bm{v},\bm{h})}}
{\sum_{\bm \sigma} e^{-E(\bm \sigma)}} = \frac{e^{-\mathcal{F}(\bm{v})}}
{\sum_{\bm{v}} e^{-\mathcal{F}(\bm{v})}},
\end{equation}
where the free energy is defined from the expression $e^{-\mathcal{F}(\bm{v})} = \sum_{\bm{h}}e^{-E(\bm{v}, \bm{h})}$.
As stated in the main text, we will consider here an RBM with no biases.
For the case of a binary hidden layer where $\bm{h}\in \{-1,1\}^{H}$, one can give a closed-form expression to it:
\begin{equation}
\mathcal{F}(\bm{v}) = \sum_{\alpha=1}^H\log\left[2\cosh\left(\frac{1}{\sqrt{K}}\sum_{i=1}^V\sum_{k=1}^K \xi_i^{(k)} \xi_{\alpha}^{(k)} v_i\right)\right].
\end{equation}
From now on, we employ roman indices for denoting the visible neurons in a pattern, and greek indices for the hidden neurons.
Therefore, for this particular case Eq.~\eqref{eq:hebbian_rule} reads \mbox{$W_{i\alpha}=\sum_{k=1}^K\xi_i^{(k)}\xi_\alpha^{(k)}/\sqrt{K}$}.

Our goal is to find the set of parameters (which we call $\theta$ for simplicity) such that $P_{\textrm{model}}$ becomes as close as possible to the $P_{\mathrm{data}}$ underlying some training dataset $\mathcal{T}$.
To compare them we employ the negative log-likelihood,
\mbox{$\mathcal{L}=-\sum_{\bm{v}^{(i)}\in\mathcal{T}} P_{\mathrm{data}}(\bm v^{(i)}) \log P_{\mathrm{model}}(\bm v^{(i)})$}.
Introducing Eq.~\eqref{prob_visible} to the previous we find that
\begin{equation}
\mathcal{L}=-\sum_{\bm{v}^{(i)}\in\mathcal{T}} P_{\mathrm{data}}(\bm v^{(i)})
\log \frac{\sum_{\bm{h}} e^{-E(\bm{v}^{(i)},\bm{h})}}{\sum_{\bm{\sigma}} e^{-E(\bm{\sigma})}}.
\end{equation}
Expanding this expression and writing it in terms of the free energy, we obtain
\begin{align}
\mathcal{L}&=-\frac{1}{|\mathcal{T}|}\sum_{\bm{v}^{(i)}\in\mathcal{T}}\log\frac{e^{-\mathcal{F}(\bm{v}^{(i)})}}{Z} \notag\\
&=\log Z - \frac{1}{|\mathcal{T}|}\sum_{\bm{v}^{(i)}\in\mathcal{T}}\log e^{-\mathcal{F}(\bm{v}^{(i)})} \notag\\
&=\log \sum_{\bm{v}} e^{-\mathcal{F}(\bm{v})} + \frac{1}{|\mathcal{T}|}\sum_{\bm{v}^{(i)}\in\mathcal{T}}\mathcal{F}(\bm{v}^{(i)}),
\end{align}
where for simplicity we have introduced the partition function $Z=\sum_{\bm v,\bm h}e^{-E(\bm v,\bm h)}$, $|\mathcal{T}|$ denotes the cardinality of $\mathcal{T}$, and we assume that $P_{\textrm{data}}(\bm v^{(i)}) = |\mathcal{T}|^{-1}\,\forall\,\bm v^{(i)}\in\mathcal{T}$ and zero otherwise.

Once the loss function is defined, we can update the weights using, e.g., the gradient descent method \mbox{$\Delta \theta = - \lambda \partial_\theta \mathcal{L}$}, which in our case means
\begin{align}\label{update_general}
\partial_\theta \mathcal{L} & =
\frac{1}{|\mathcal{T}|}\sum_{\bm{v}^{(i)}\in\mathcal{T}}\partial_\theta\mathcal{F}(\bm{v}^{(i)})
-\frac{1}{Z}\sum_{\bm{v}} e^{-\mathcal{F}(\bm{v})}\partial_\theta \mathcal{F}(\bm{v})\nonumber\\
& = \frac{1}{|\mathcal{T}|}\sum_{\bm{v}^{(i)}\in\mathcal{T}}\partial_\theta\mathcal{F}(\bm{v}^{(i)})
-\sum_{\bm{v}} P_{\mathrm{model}}(\bm v)\partial_\theta \mathcal{F}(\bm{v}).
\end{align}

One can distinguish clearly here the \textit{positive} and \textit{negative} phases.
The positive phase is the first term, evaluated only on the instances of the training set, while the negative phase is the negative term, that is evaluated on every possible configuration of the visible nodes.

In the case of standard RBMs, the ultimate parameter that one desires to fix are the weights $W_{i\alpha}$.
For these, the derivative of the free energy function (this is, the function $\mathrm{get}\_\mathrm{phase}()$ in Algorithm~\ref{alg:rapid}) is
\begin{equation}
\label{d_fe_J}
\frac{\partial\mathcal{F}(\bm{v})}{\partial W_{i\alpha}} = -v_i\tanh\left(\sum_j W_{j\alpha}v_j\right),
\end{equation}
where we have assumed the requirements of the models in the main text, namely that we have spin variables (i.e., $\sigma_i=\pm 1$) and that all biases are zero.

On the other hand, when considering an RA-RBM, the ultimate parameters to be determined are the auxiliary patterns $\bm\xi^{(k)}$, with which the weights are later computed by using Eq.~(5).
In this case the function $\mathrm{get}\_\mathrm{phase}()$ is the gradient of the free energy with respect to the individual pattern neurons $\xi_j^{(k)}$, which evaluates to
\begin{subequations}\label{eq:updaterule}
    \begin{align}
        \frac{\partial\mathcal{F}(\bm{v})}{\partial\xi^{(k)}_i} & = \frac{1}{\sqrt{K}}v_i\!\sum_{\alpha=1}^H\xi^{(k)}_{\alpha}\tanh\!\left(\!\frac{1}{\sqrt{K}}\sum_{m=1}^K\sum_{j=1}^V\xi_{j}^{(m)}\xi_\alpha^{(m)} v_j\right) \nonumber \\
        &= \frac{1}{\sqrt{K}}v_i\sum_{\alpha=1}^H\xi^{(k)}_{\alpha}\tanh\left(\sum_{j=1}^V W_{j\alpha}v_j\right),
    \end{align}
    \begin{align}
        \frac{\partial\mathcal{F}(\bm{v})}{\partial\xi^{(k)}_\alpha} & =  \!\frac{1}{\sqrt{K}}\!\!\left(\sum_{i=1}^V v_i\xi^{(k)}_{i}\!\!\right)\!\tanh\!\!\left(\!\frac{1}{\sqrt{K}}\!\!\sum_{m=1}^K\sum_{j=1}^V\xi_{j}^{(m)}\xi_\alpha^{(m)} v_j\!\right) \nonumber \\
        & =  \frac{1}{\sqrt{K}}\left(\sum_{i=1}^V v_i\xi^{(k)}_{i}\right)\tanh\left(\sum_{j=1}^V W_{j\alpha} v_j\right),
    \end{align}
\end{subequations}
depending on whether the neuron is visible or hidden, respectively.
\section{From continuous updates to discrete patterns}\label{app:discrete}
\setcounter{equation}{0}
After the update of the patterns according to Eqs.~(4) and~\eqref{eq:updaterule}, the values of the neurons will be continuous, losing its meaning as spin configurations, and with it the guarantee that they represent low-energy configurations of the associated Ising system.
In the following we describe three methods to bring the continuous-valued, updated parameters $\xi_i^{(k)}$ back into discrete, real spin configurations:
\begin{enumerate}
\item \underline{Sign discretization:} The first method amounts to simply substitute the value of each of the continuous variables by its sign, i.e.,
\begin{equation}
    \xi^{(k)}_i \leftarrow \mbox{sign}(\xi^{(k)}_i).
\end{equation}
This not only ensures that the auxiliary neurons are binary, but also acts as a regularizer, avoiding divergences.

\item \underline{Value restriction:} When training in more complex datasets, the expressivity of the models can be enhanced by considering that the auxiliary neurons are continuous.
In such case, we no longer discretize them.
In order to prevent the divergence of the weights, we restrict $\xi_i^{(k)} \in [-x,x] \ \forall \ i,k$. The value of $x$ is arbitrary.
In the examples shown in this work we choose $x\,{=}\,1$.
However, for other values considered, we observe similar results in terms of training quality.

\item \underline{Gibbs sampling:} For small learning rates, the updated patterns remain close to spin configurations.
Thus, a way of obtaining spin configurations in the low-energy sector is performing Gibbs steps, taking as initial seeds the values of the continuous patterns.
This is, one would perform:
\begin{subequations}
    \begin{align}
        \xi^{(k)}_{i\in V}\sim p\left(\xi^{(k)}_i=1\middle|\bm\xi^{(k)}_h\right)=\sigma\left(2\sum_{\alpha=1}^H\xi^{(k)}_\alpha W_{i\alpha}\right), \\
        \xi^{(k)}_{\alpha\in H}\sim p\left(\xi^{(k)}_\alpha=1\middle|\bm\xi^{(k)}_v\right)=\sigma\left(2\sum_{i=1}^V\xi^{(k)}_iW_{i\alpha}\right),
    \end{align}
\end{subequations}
where $\sigma(x)=(1+e^{-x})^{-1}$ is the sigmoid function.
This procedure not only transforms the patterns back into spin configurations, but also forces them to lie in the low-energy spectrum of the Ising model and inserts mixing, which can be beneficial in the late stages of training.
\end{enumerate}

In procedures 1) and 2) it is crucial to choose when to perform either the discretization or the restriction in the given range.
For the former, discretizing too often may result in the erasure of the information learnt by the model, as the cumulant of the updates before the discretization may not be large enough to change the sign of a given $\xi$.
Not discretizing often enough may result on a similar phenomenon.
For example, if the first updates of a $\xi^{(k)}_i=1$ are positive, subsequent negative updates will have no effect in $\xi^{(k)}_i$, as its value may be very far from zero.
This method was applied to the BAS examples in Section~\ref{sec:results}, both the $4\,{\times}\,4$ and $12\,{\times}\,12$ datasets, by discretizing the patterns after every epoch of training (this is, after the end of every pass of the full dataset).
Note that an alternative solution may be to employ a learning rate large enough so as to permit an appropriate size of the cumulants.
For the latter, i.e., the restriction of $\xi_i^{(k)} \in [-x,x]$, a similar approach holds.
However, given that the effect of such procedure on the value of the patterns is not as dramatic as in the previous case, it can be applied much more often.
We show the validity of this procedure in the MNIST example.
There, the value of the patterns is checked and bounded after every update.

The frequency with which procedure 3) is applied can also be chosen at will.
However, it is very natural to perform it at every training step.
Note that in such a case one would have a variant of CD, and therefore speedups in learning when compared to standard BMs would only be attributed to regularizing the axons through Eq.~\eqref{eq:hebbian_rule}.

\section{Details on benchmarking through Hamming distance}\label{app:task}
\setcounter{equation}{0}
Given an image A, we fix the top row of pixels (perpendicular to the direction of the bars), and set the remaining pixels to have the value 0.
We perform Gibbs sampling, allowing for the rest of the visible neurons to be updated.
Then, the Hamming distance between the model-generated image B and the given one A is calculated.
We normalize such distance by dividing over the number of pixels of the image, this is, the number of visible neurons $V$.
In this way, for a random reconstruction where a pixel has a probability of $0.5$ of coinciding with the desired one, the average Hamming distance with the original image will be also $0.5$.
As an example of this procedure, in Fig.~\ref{fig:corruption} we show the corrupted version of the set of $4\,{\times}\,4$ bars images in the test set in the experiments in Section~\ref{sec:results:smallbas}.

\renewcommand\thefigure{\thesection.\arabic{figure}}
\setcounter{figure}{0}
\begin{figure}[h]
    \centering
    \includegraphics[width=.9\columnwidth]{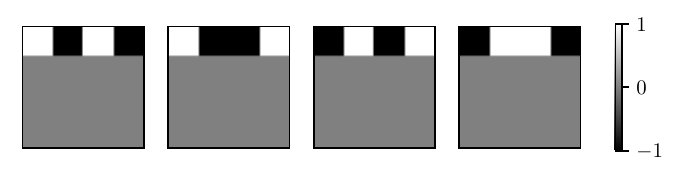}
	\caption{\textbf{Corrupted images for benchmarking:}  Example of image corruption in the $4\,{\times}\,4$ Bars test set. For addressing reconstruction, the visible neurons are initialized in such configurations and, after a number of Gibbs steps, the outputs are compared to the original images via their Hamming distance. The same approach is followed for assessing training in the $12\,{\times}\,12$ BAS dataset in Section~\ref{sec:results:complex}.
	}
	\label{fig:corruption}
\end{figure}

\section{Computational cost of CD vs. PID}\label{app:cost}
\setcounter{equation}{0}
In this appendix we analytically calculate the computational cost of training RBMs following both PID and Contrastive Divergence.
The main difference in complexity is the calculation of the negative phase.
In the case of PID, computing this term is trivial as it just involves averages over the auxiliary patterns.
However, one needs to take into account the cost of calculating the weights after each update following Eq.~\eqref{eq:hebbian_rule}.
This implies doing a sum of $K$ elements for each weight $W_{i\alpha}$.
As we have $V H$ weights, the cost of this operation is $\mathcal{O}(KVH)$.

In the case of CD and its variants, the most basic algorithm consists on doing $k$ Gibbs steps from a batch of $f$ initial visible configurations.
From here, one calculates the activation probability of each hidden neuron $h_\alpha$ as (note that, for simplicity, we depict a standard RBM with nonzero bias and with neurons taking values $\{0,1\}$)
\begin{equation}
\label{eq:hid_prob}
    p(h_\alpha=1|\bm{v})=\sigma\left(\sum_i v_i W_{i\alpha} + b_\alpha\right).
\end{equation}
This calculation has a computational cost of $\mathcal{O}(V H)$. Note that here we consider the general case where the biases $b_j$ can take nonzero values.
To complete a Gibbs step, one needs to calculate the value of the visible layer given the hidden vector obtained from Eq.~\eqref{eq:hid_prob} by
\begin{equation}
\label{eq:vis_prob}
    p(v_i=1|\bm{h})=\sigma\left(\sum_\alpha h_\alpha W_{i\alpha} + c_i\right),
\end{equation}
which again has a computational cost of $\mathcal{O}(H V)$.
Summing both contributions and taking into account that these procedure is performed $k$ times to approximate unbiased sampling for each initial configuration, the total complexity of CD scales as $\mathcal{O}(2kfHV)$.

\section{Regularized Axons and Hopfield patterns}\label{app:hopfield}
In this appendix we expand the discussion of the Regularized Axons that this work proposes from perspective of the correspondence between BMs and Hopfield models.
Generally one cannot simply reduce BMs with binary units to the Hopfield model by marginalization over hidden units.
First, in case of deep BMs such marginalization is an NP-complete problem.
Second, even in the case of RBMs where the marginalization over hidden units is computable, the resulting model contains interactions of any order (i.e., two-body, three-body, four-body, etc.) between the visible units.
This is in stark contrast to the Hopfield model.
However, one can consider non-standard BMs which would be reduced exactly to the Hopfield model upon marginalization over hidden units. In Ref.~\cite{2012Barra} the authors proposed a Hybrid Boltzmann Machine (HBM), with continuous hidden neurons and binary visible neurons, with this property.
The corresponding model analogous to an RBM can be shown to be equivalent to the Hopfield model with weights given by~\cite{2012Barra}:
\begin{equation}
  \label{eq:hybridBM}
  J_{ij}=\sum_{h} W_{ih}W_{hj},
\end{equation}
where $W_{ih}$ are the weights of the Restricted HBM and summation is over the index corresponding to the hidden units.
When one considers the case of binary weights, the $W_{ih}$ can be interpreted as patterns that give rise to Hebbian weights in an equivalent Hopfield model.
In such case, the number of patterns is equal to the number of hidden units in the Restricted HBM.
This imposes an interesting constraint on the number of hidden units in the Restricted HBM with binary weights by the capacity of retrievable patterns in the equivalent Hopfield network.

In order to highlight the difference of working with patterns in BMs and in Hopfield models, let us now regularize the weights in the HBM of Ref.~\cite{2012Barra} according to Eq.~\eqref{eq:hebbian_rule}.
Note that, upon application of Eq.~\eqref{eq:hebbian_rule}, the weights in the resulting RA-HBM model are no longer binary (even approximately) and cannot be interpreted as Hebbian patterns in the resultant Hopfield network. The number of patterns $K$ in the RA-HBM is now an additional parameter to the number of hidden units. Eq.~\eqref{eq:hybridBM} can now be recast as:
\begin{equation}
  \label{eq:hybridBM2}
  J_{ij}=\frac{1}{K}\sum_{k,k'}\xi_i^{(k)}\xi_j^{(k')}M_{k,k'},
\end{equation}
where $M_{k,k'}=\sum_h \xi_h^{(k)}\xi_h^{(k')}$.
This means that the Hebbian form of the weights in RA-HBMs does not correspond to the Hebbian form of weights in the resultant Hopfield model.
Note that the Hebbian form of weights in Hopfield models is the simplest construction of retrievable configurations, but it is by no means optimal.
The maximum capacity for Hebbian patterns is $0.14 N$, where $N$ is the number of visible units~\cite{1985Amit}.
However, the maximum theoretical capacity of retrievable configurations when the weights are unrestricted is $2N$~\cite{Cover1965,Gardner_1988}.
It is an interesting direction of future work to see whether the form of weights given by Eq.~\eqref{eq:hybridBM2} would allow to reach higher capacities of retrievable configurations in the Hopfield model.

Given the argument above on the differences between RA models and Hopfield models, other interesting, non-standard Hopfield models with trainable Hebbian patterns like those in Refs.~\cite{weigt1,weigt2} are analogously not directly related to the RA approach proposed in this work.
\end{document}